\documentclass[lettersize,journal]{IEEEtran}
\usepackage{amsmath,amsfonts}
\usepackage{algorithmic}
\usepackage{array}
\usepackage[caption=false,font=normalsize,labelfont=sf,textfont=sf]{subfig}
\usepackage{textcomp}
\usepackage{stfloats}
\usepackage{url}
\usepackage{verbatim}
\usepackage{graphicx}
\usepackage{cite}

\usepackage{enumerate}
\usepackage{amsfonts,mathrsfs,amssymb,bm,mathtools,mathabx,bbm}
\usepackage{multirow,mdframed}
\usepackage[ruled,linesnumbered,noresetcount,vlined]{algorithm2e}
\usepackage{booktabs}
\usepackage{enumitem}
\usepackage{todonotes}

\DeclareMathOperator*{\argmin}{argmin}

\newcommand{\cO}{\mathcal{O}}
\newcommand{\bc}{\bm{c}}
\newcommand{\bx}{\bm{x}}

\newcommand{\bz}{\bm{z}}

\newcommand{\cX}{{\cal X}}

\newcommand{\RR}{\mathbb{R}}

\usepackage{float} 
\floatstyle{ruled}
\newfloat{model}{thp}{lop}
\floatname{model}{Model}

\begin{document}

\title{Learning Regionally Decentralized\\ AC Optimal Power Flows with ADMM}

\author{Terrence W.K.~Mak,
        Minas Chatzos, 
		Mathieu Tanneau,
		and~Pascal Van Hentenryck
    \thanks{The authors are affiliated with the H. Milton Stewart School of Industrial and Systems Engineering, 
	Georgia Institute of Technology, Atlanta, GA 30332. 
	e-mail contacts: \{wmak,minas\}@gatech.edu, \{mathieu.tanneau,pvh\}@isye.gatech.edu.
}}

\markboth{}%
{}


\maketitle

\begin{abstract}
One potential future for the next generation of smart grids is the use
of decentralized optimization algorithms and secured communications
for coordinating renewable generation (e.g., wind/solar), dispatchable
devices (e.g., coal/gas/nuclear generations), demand response, battery
\& storage facilities, and topology optimization. The Alternating
Direction Method of Multipliers (ADMM) has been widely used in the
community to address such decentralized optimization problems and, in
particular, the AC Optimal Power Flow (AC-OPF). This paper studies how
machine learning may help in speeding up the convergence of ADMM for
solving AC-OPF. It proposes a novel {\em decentralized}
machine-learning approach, namely ML-ADMM, where each agent uses deep
learning to learn the consensus parameters on the coupling
branches. The paper also explores the idea of learning only from ADMM
runs that exhibit high-quality convergence properties, and proposes
filtering mechanisms to select these runs. Experimental results on
test cases based on the French system demonstrate the potential of the
approach in speeding up the convergence of ADMM significantly.
\end{abstract}


\textcolor{red}{%
}

\section{Introduction}

One potential future for the next generation of smart grids \cite{Fang12Smart} is the use of decentralized optimization algorithms and secured communications for coordinating renewable generation (e.g., wind/solar), dispatchable devices (e.g., coal/gas/nuclear
generations), demand response, battery \& storage facilities.
In particular, system operators will need to reliably and efficiently solve AC Optimal Power Flow (AC-OPF) problems in a decentralized fashion.
This optimization problem finds the most economical generation dispatch that meets the load, while also satisfying the physical and engineering constraints of the underlying power grid.
It is therefore a fundamental tool for balancing generation and load rapidly, without sacrificing economic efficiency.
Nevertheless, its resolution in a decentralized fashion remains challenging, especially for industry-size networks that comprise thousands of buses.

The alternating direction method of multipliers (ADMM)
\cite{Boyd2011_ADMM} is widely used by the power systems community to
solve decentralized optimization problems, especially OPF
problems~\cite{molzahn2017survey}.  In particular, ADMM has been
successfully applied to convex relaxations and/or approximations of
AC-OPF, e.g., the popular DC approximation \cite{Wang17fully}, for
which it enjoys strong theoretical guarantees.  Furthermore, ADMM
schemes with convergence guarantees have been proposed recently (e.g.,
\cite{sun21two}), broadening the scope of application of this
decentralized optimization technology. However, most ADMM variants
used for AC-OPF, including those with convergence guarantees, require
significant tuning of their parameters to ensure numerical stability
and convergence in practice \cite{molzahn2017survey}.

\begin{figure}[!t]
    \centering
    \includegraphics[width=0.95\linewidth]{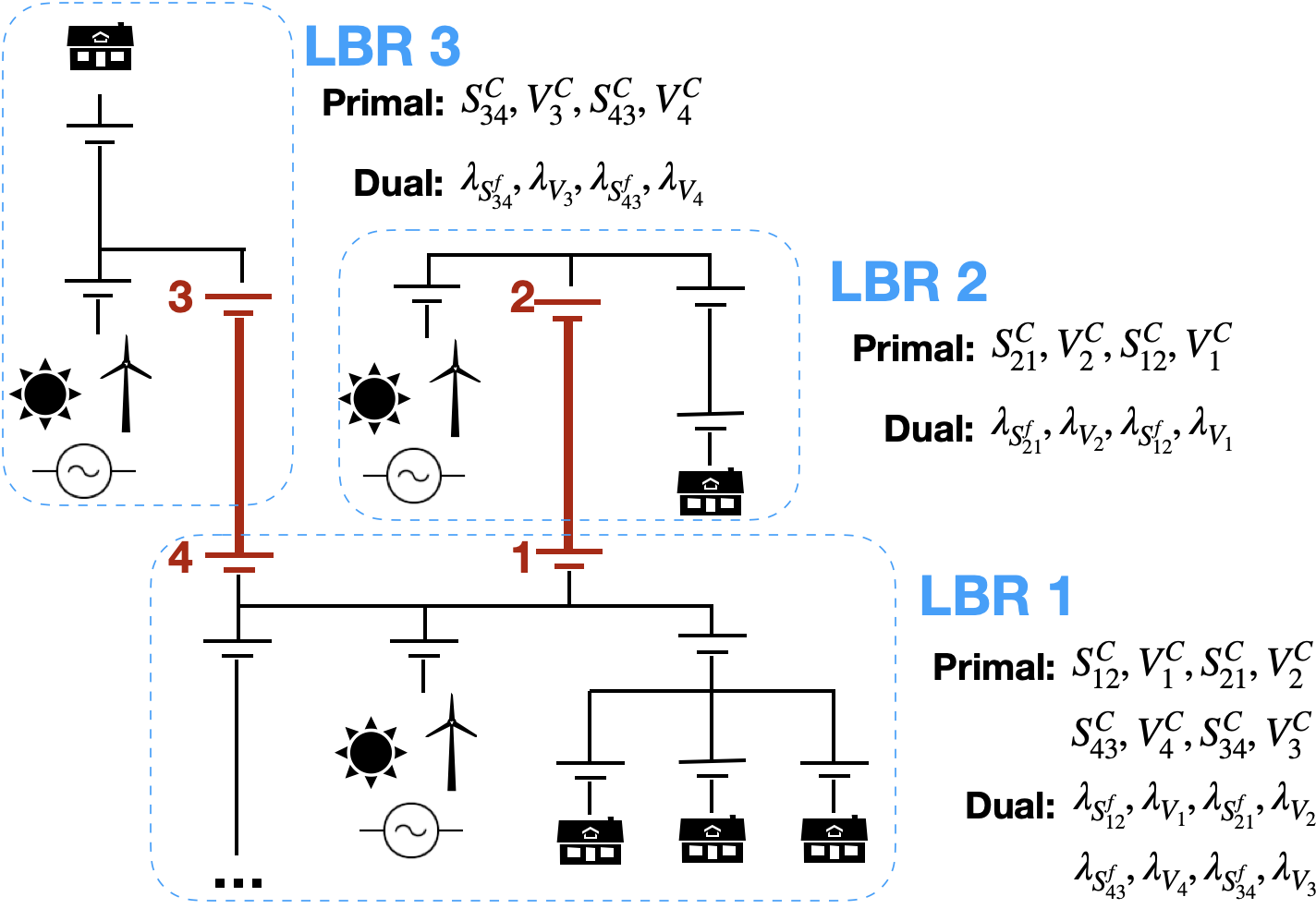}
    \caption{Power Grid Example with 3 Load Balancing Regions. Consensus constraints are formulated on the two coupling lines $(1, 2)$ and $(3, 4)$ and the corresponding buses.}
    \label{fig:example1}
\end{figure}

In that context, this paper proposes the use of machine learning (ML) techniques to enhance the practical behavior of ADMM for solving AC-OPF problems in a decentralized fashion.
The paper leverages the fact that ADMM is an iterative process that uses dual (Lagrange) multipliers to drive separate agents towards achieving a consensus \cite{Boyd2011_ADMM}.
This perspective is illustrated in Figure \ref{fig:example1}, which depicts a power grid composed of 3 regions: the regions are coupled through lines $(1, 2)$ and $(3, 4)$ for which a consensus much be reached.
Building on this observation, the paper proposes ML-ADMM, which uses ML to learn a close-to-optimal primal-dual solution that is used to warm-start the ADMM algorithm.
Specifically, the paper makes the following contributions:
\begin{enumerate}
    \item it proposes to learn both primal \emph{and dual} consensus variables, in contrast with other works that only consider primal information;
    \item it introduces a novel decentralized machine learning approach for data collection, training and inference;
    \item it proposes novel data-filtering techniques to identify high-quality training data, thereby improving training and learning accuracy;
    \item it reports computational results on real, industry-scale systems from the French transmission grid;
    \item it demonstrates the applicability of the methodology on two classes of ADMM schemes, one of which has strong convergence guarantees;
    \item The numerical experiments show that ML-ADMM
obtains solutions of similar quality as the original ADMM schemes in as little as 1/6 of the iterations.
\end{enumerate}

It is important to note that the proposed ML-ADMM framework is not
restricted to AC-OPF, and can be applied to other optimization
problems.  In addition, because ML-ADMM executes the ADMM algorithm
from a high-quality starting point, it enjoys the same theoretical
convergence properties.


The rest of the paper is organized as follows.
Sections \ref{sec:related} and \ref{sec:background} present the related work and background material.
Section \ref{sec:distributedOPFs} introduces the Regionally Decentralized AC-OPFs and the ADMM formulations.  
Section \ref{sec:distributedlearning} and \ref{sec:data_filtering} present the decentralized learning models and data filtering procedures for learning the interconnection parameters. 
Section \ref{sec:evaluation} reports the experimental results.
Section \ref{sec:conclusion} concludes the paper.

\vspace{1cm}

\section{Related Work}
\label{sec:related}
With the introduction of Smart Grid, there has been a growing interest in applying ADMM on optimal power flow applications, primarily due to its distributive nature and privacy features~\cite{sun13fully, erseghe14distributed, Magnusson15distributed, mhanna18component, mhanna19adaptive}.
Even though general ADMM formulations on nonlinear AC-OPF may not always converge, recent work~\cite{sun21two} shows  that convergence can be achieved via reformulations under mild assumptions. This paper complements these convergence results by showing how warm-starting
the ADMM formulations through machine learning can bring substantial speedups in practice.

The application of machine learning  
to optimal power flows has been widely studied in recent years.
A recent line of research has focused on how to predict
centralized AC-OPFs solutions directly using Deep Neural Networks (DNN)~(e.g.,
\cite{fioretto2020predicting,yan20real,pan19deepopf,OPFLearningTutorial21}).
Once a neural network is trained, solution predictions can be computed 
with a single forward pass in milliseconds.
Recent work~\cite{chatzos21spatial} has also
shown that deep learning can be spatially decomposed in a similar
fashion.
A wide variety of approaches have also been proposed beyond 
predicting AC-OPF solutions.
These approaches include learning
the active set of constraints~\cite{misra2019learning,xavier2019learning,deka2019learning,hasan2020hybrid,robson2020learning},
imitating the Newton-Raphson algorithm~\cite{baker2020learningboosted},
learning warm starting points for speeding-up 
the optimization process~\cite{baker2019learning,chen2020hotstarting},
and predicting optimal dispatch decisions~\cite{pan19deepopf,pan2020deepopf,zamzam2020learning}. 
Applying machine learning techniques to decentralized OPF
problems has also been studied recently~\cite{Biagioni22learning}.
Other related works
explore formal guarantees for neural networks when learning OPF problems ~\cite{venzke2020verification,venzke2020learning},
and extend the learning methodologies to security-constrained OPF problems ~\cite{pan2020deepopf2,velloso2020combining}.
Reinforcement-learning approaches for OPF problems have also been proposed (e.g.,~\cite{zhou20datadriven,yan20realtime,woo20realtime,Sanseverino16multiagent}) and  primarily focus on tackling real-time issues. This paper continues the line of work~\cite{fioretto2020predicting,chatzos2020highfidelity,chatzos21spatial,chen2021learning} 
in using deep learning to predict AC-OPF solutions directly, while  
integrating practices/constraints found in U.S. energy markets. 
The work differs from existing work on decentralized OPFs (e.g., \cite{Biagioni22learning}) in three ways:
\begin{enumerate}
\item it focuses on predicting the flows and voltage on coupling branches, governed largely by transactions between regional load balancing zones/authorities (for the inter-regional exchange markets) instead of learning the decentralized algorithm itself, e.g., learning the search directions/heuristics;
\item the learning procedure is decentralized by nature and
each agent can train in parallel and independently, maintaining privacy \& region/agent neutrality; and
\item the predictions are not necessarily tied to any specific decentralized algorithm, and can be seen as predicting transactions in an exchange market.
\end{enumerate}

\begin{table*}[!t]
\caption{Notation \& Symbols \label{tab:notation}}
\begin{center}
	{
	\begin{tabular}{|l l | l l|}
    \toprule
        Symbol                         & Description                           & Symbol                & Description \\
        \midrule
		{$\bm{\mathcal N} = (N,E)$}  	& Power grid 				            & $j$                   & Imaginary unit  \\
		{$N$}  	                        & Set of buses  			            & $V_i = v_i \angle \theta_i$ & Bus voltages of bus $i$\\
		{$E$}       			        & Set of branches 			            & $S^f_{ij} = p^f_{ij} + j q^f_{ij}$ & Line power flow of branch $(i,j)$\\
	$E^R = \{(j,i) : (i,j) \in E \}$ & Set of branches in reverse direction   & $Y_{ij} = g_{ij} + j b_{ij}$       & Line Admittance of branch $(i,j)$\\
		{$G$}		                    & Set of generators  		            & $S^g_i = p^g_i + j q^g_i$ & Generation dispatch of generator $i$\\
        {$D$}		                    & Set of load demands  		            & $S^d_i = p^d_i + j q^d_i$ & Load demand of load $i$\\
        $s \in N$                       & Reference bus / slack bus             & $M_i$                     & Market cost function for generator $i$ \\
        \midrule
        $K$                             & Set of regions ($k \in K$)            & {$N_k \subseteq N$} & Set of local buses at region $k$\\ 
		$E_k \subseteq E$               & Set of local branches at region $k$   & {$R_k \subseteq E$ }& Set of inter-regional branches at region $k$\\
		{$G_k \subseteq G$}             & Set of local generators at region $k$ & $D_k \subseteq D$& Set of local demands at region $k$\\
		\multirow{ 2}{*}{$N^B_k \subseteq N_k$}       & Set of border buses at region $k$     & \multirow{ 2}{*}{$N^N_k \not\subseteq N_k$} & Set of neighbouring buses at neighbouring regions\\
		                                & connecting to other regions           &           &  connecting to region $k$\\
		\midrule
        $x^*$                           & Complex conjugate of quantity $x$     & $\overline{x}, \underline{x}$  & Upper and lower bound of quantity $x$     \\
        $\widehat{x}$                       & Prediction/Forecast of quantity $x$   & $\Re(x), \Im(x)$ & Real and imaginary component of complex quantity $x$\\
        ${x}^{C}$                       & Consensus of quantity $x$             & $x[k]$      & Projection of quantities $x$ to region/area $k$ \\
        $\rho$                          & penalty term (ADMM)                   & $\lambda$   & Lagrangian multipliers (ADMM)\\
	\bottomrule
  	\end{tabular}
  	}
	\end{center}
  	
\end{table*}
\section{Background}
\label{sec:background}
This section presents background materials for the rest of the paper.
Table~\ref{tab:notation} presents the common notations and symbols.

\subsection{AC Optimal Power Flow}
The AC Optimal Power Flow (OPF) determines the most economical
generation dispatch balancing the load and generation in a power
grid. 
\begin{model}[!t]
	{\small
	\caption{AC Optimal Power Flow: $P_{AC}$}
	\label{model:ac_opf}
	\vspace{-6pt}
	\begin{align}
        \mbox{\bf input:} \;\; & \bm{S^d} = ( S^d_i: i\in N) \nonumber \\        
		\mbox{\bf variables:} \;\;
		& \bm{S^g} =(S^g_i: i \in N),  \bm{V} = (V_i: i \in N) \nonumber \\
		{ } &   \bm{S^f} = (S^f_{ij} : \forall(i,j)\in E \cup E^R) \nonumber \\  		  
		\mbox{\bf minimize:} \;\;
		& {\cO}(\bm{S^g}) = \sum_{i \in N} M_{i}(\Re(S^g_i)) \label{ac_obj} \\
		\mbox{\bf subject to:} \;\; 
		& \theta_{s} = 0, \label{eq:ac_0} \\
		& \underline{v}_i \leq v_i \leq \overline{v}_i  		\;\; \forall i \in N \label{eq:ac_1} \\
		& \overline{S}^g_i \leq S^g_i \leq \underline{S}^g_i \;\; \forall i \in N \label{eq:ac_3}  \\
		& |S^f_{ij}| \leq \overline{S}_{ij} 					\;\; \forall (i,j) \in E \cup E^R \label{eq:ac_4}  \\
		& S^g_i - {S}^d_i = \textstyle\sum_{(i,j)\in E \cup E^R} S^f_{ij} \;\; \forall i\in N \label{eq:ac_5}  \\ 
		& S^f_{ij} = {Y}^*_{ij} |V_i|^2 - {Y}^*_{ij} V_i V^*_j 			 \;\; \forall (i,j)\in E \cup E^R \label{eq:ac_6}
	\end{align}
	}
	\vspace{-10pt}
\end{model}
Model~\ref{model:ac_opf} presents 
an AC OPF
formulation (centralized model), with variables and parameters in the complex domain
for clarity and compactness.
For simplicity, the presentation omits the equations for transformers, 
phase shifters, circuit breakers/switches, and fixed/switched bus shunts.
All omitted devices are considered and implemented in the 
experimental evaluation.
The objective function ${\cO}(\bm{S^g})$ captures the
cost of the generator dispatch, with $\bm{S^g}$ denoting the vector of
generator dispatch values $(S^g_i \:|\: i \in N)$.  Constraint
\eqref{eq:ac_0} sets the voltage angle of the reference/slack bus $s
\in N$ to zero to eliminate numerical symmetries.  Constraint
\eqref{eq:ac_1} bounds the voltage magnitudes.
Constraint \eqref{eq:ac_3} enforces the generator output $S^g_i$ to
stay within its limits.  Constraint
\eqref{eq:ac_4} imposes the line flow limits on all the
line flow variables $S_{ij}$.  Constraint \eqref{eq:ac_5} captures
Kirchhoff's Current Law enforcing the flow balance of generations
$S^g_i$, loads $S^d_i$, and branch flows $S_{ij}$ across every node.
Finally, constraint \eqref{eq:ac_6} captures Ohm's Law describing the
AC power flow $S_{ij}$ across lines/transformers.

\subsection{Alternating Direction of Multipliers Method (ADMM)}
\label{sec:admm}
ADMM~\cite{Boyd2011_ADMM} is a widely used decentralized algorithm solving decentralized optimization problems with coupling constraints.
Consider an optimization problem with two agents/parties:
\begin{align}
\min_{\bx_1, \bx_2}\quad & f_1(\bx_1) + f_2(\bx_2) \nonumber\\
\mbox{ s.t. } \quad & \bx_1 \in \cX_1, \bx_2 \in \cX_2, \nonumber \\
                    & A\bx_1 + B\bx_2 = \bc, \label{math:admm}
\end{align}
where $\cX_1 \subseteq \RR^n$ and $\cX_2 \subseteq \RR^m$ are two disjoint feasible space for two independent local optimization problems, $\bx_1 \in \cX_1 \subseteq \RR^n$ and $\bx_2 \in \cX_2 \subseteq \RR^m$ denote feasible variable vectors owned by two distinct groups of agents, and $A \bx_1 + B \bx_2 = \bc$ describes the set of $l$ \emph{coupling constraints} between the two groups of agents with $A \in \mathbb{R}^{\ell \times n}$, $B \in \mathbb{R}^{\ell \times m}$, and $\bc \in \mathbb{R}^{\ell}$. 
The functions $f_1$ and $f_2$ denote the objectives over $\bx_1$ and $\bx_2$, respectively.
They are commonly assumed to be convex.

Problem (\ref{math:admm}) is often reformulated and simplifed by
introducing consensus parameters explicitly, leading to the
consensus formulation~\cite{mhanna19adaptive}.  Let $\bx_1^C, \bx_2^C$
to be the consensus for $\bx_1, \bx_2$.  The consensus formulation of
(\ref{math:admm}) for agent 1 is:
\begin{align}
\min_{\bx_1}\quad & f_1(\bx_1) \nonumber\\
\mbox{ s.t. } \quad & \bx_1 \in \cX_1,  A\bx_1 = \bc - B\bx_2, \nonumber\\
\mbox{ where }\quad & \bx_2 = \bx_2^C \label{math:consensus}
\end{align}
The consensus formulation for agent 2 is similar. The augmented Lagrange function $L^1_{\rho}(\bx_2^C, \bm{\lambda}_2)$ of (\ref{math:consensus}) for agent 1 is:
\begin{align*}
\min &f_1(\bx_1) + \bm{\lambda}_2^\intercal \bx_2 + \frac{\rho}{2} \lVert \bx_2 - \bx_2^C \rVert^2_2
\\
\mbox{ s.t. } \quad & \bx_1 \in \cX_1,  A\bx_1 = \bc - B\bx_2, \nonumber
\end{align*}
where $\bm{\lambda}_2$ is a vector of \emph{Lagrangian multipliers} for $\bx_2$ in the view of agent 1,
with $\rho > 0$ representing the penalty parameter. 
Similarly, the augmented Lagrange function $L^2_{\rho}(\bx^C_1, \bm{\lambda}_1)$ of (\ref{math:consensus}) for agent 2 is:
\begin{align*}
\min &f_2(\bx_2) + \bm{\lambda}_1^\intercal \bx_1 + \frac{\rho}{2} \lVert \bx_1 - \bx_1^C \rVert^2_2
\\
\mbox{ s.t. } \quad & \bx_2 \in \cX_2,  B\bx_2 = \bc - A\bx_1 \nonumber
\end{align*}
where $\bm{\lambda}_1$ is a vector of \emph{Lagrangian multipliers} for $\bx_1$ in the view of agent 2.

Given a solution tuple $(\bx_1^k, \bx_2^k)$ and the Lagrangian multipliers $(\bm{\lambda}_1^k, \bm{\lambda}_2^k)$ at iteration $k$, 
ADMM proceeds to the next iteration, $k + 1$, 
as follows:
\begin{align}
\bx_1^{k+1} &= \argmin_{\bx_1} L^1_{\rho}(\bx_2^k, \bm{\lambda}_2^k) \label{math:admm_stepx} \\
\bx_2^{k+1} &= \argmin_{\bx_2} L^2_{\rho}(\bx_1^{k+1}, \bm{\lambda}_1^k) \label{math:admm_stepy}\\
\bm{\lambda}_1^{k+1} &= \bm{\lambda}_1^{k} + \rho (\bx^{k+1}_1 - \bx^k_1) \mbox{, and }\nonumber \\
\bm{\lambda}_2^{k+1} &= \bm{\lambda}_2^{k} + \rho (\bx^{k+1}_2 - \bx^k_2)  \label{math:admm_lambda}
\end{align}
The algorithm terminates when a desired condition (e.g., an iteration limit or a convergence factor) is reached.
The quality of the solution at iteration $k$ can be measured by the primal infeasibility (residue) vector~\cite{mhanna19adaptive}
\begin{align}
\bm{r}_{p}^k = A\bx_1^k + B\bx_2^k - \bc,
\end{align}
indicating the distance to a primal feasible solution, and 
the dual infeasibility (residue) vector~\cite{mhanna19adaptive}
\begin{align}
\bm{r}_{d}^k = \rho A^{T} B (\bx_2^k - \bx_2^{k - 1}),
\end{align}
indicating the distance from the previous local minima.  When both
infeasibility vectors are zero, ADMM has converged to a (local)
optimal and feasible solution.

\subsection{Two-level ADMM}
The two-level ADMM~\cite{sun21two} reformulates \eqref{math:admm} by introducing slack variables $\bz$  for the set of coupling constraints:
\begin{align}
\min_{\bx_1, \bx_2}\quad & f_1(\bx_1) + f_2(\bx_2) \nonumber\\
\mbox{ s.t. } \quad & \bx_1 \in \cX_1, \bx_2 \in \cX_2, \nonumber \\
                    & A\bx_1 + B\bx_2 + \bz = \bc, \label{math:two-level-admm} \\
                    & \bz = 0. \nonumber 
\end{align}
The scheme ensures that the slack eventually converges to zero (i.e.,
$\bz = 0$). The augmented Lagrange function for agent 1 then becomes
\begin{align*}
\min & \; f_1(\bx_1) + \Lambda^\intercal \bz + 
\frac{\beta}{2} \lVert \bz \rVert^2_2 + 
\bm{\lambda}_2^\intercal \bx_2 + \frac{\rho}{2} \lVert \bx_2 - \bx_2^C + \bz \rVert^2_2
\\
\mbox{ s.t. } \quad & \bx_1 \in \cX_1,  A\bx_1 = \bc - B\bx_2, \nonumber
\end{align*}
where $\Lambda$ is a vector of additional Lagrangian multipliers for
the slack variables, and $\beta > 0$ is their the associated penalty
parameter.

The two-level ADMM method uses a two-level nested loop to search and
update the two different sets of Lagrangian multipliers ($\lambda$ and
$\Lambda$).  The inner loop updates $\lambda$ together with the slack
parameters $\bm{z}$ based on a fixed $\Lambda$, while the outer loops
updates $\Lambda$ and the associated penalty $\beta$ based on the
latest solution computed from inner loop.  The inner and outer loops
can also be implemented as a single loop where the updates by the
outer loop are performed only if the search meets certain
predetermined criteria $\eta(\cdot)$.

\subsection{Deep Learning Neural Network (DNN)}
Deep Neural Networks are a learning framework composed of a
sequence of layers, with each layer typically taking as inputs the results of the
previous layer (e.g., \cite{lecun2015deep}). 
Commonly used Feed Forward Neural Networks (FNNs) are
DNNs where the layers are fully connected.
The function connecting the layers, from $\mathbb{R}^n$ to $\mathbb{R}^m$ is given by: 
$$
\bm{y} = \pi(\bm{W} \bm{x} + \bm{b}),
$$ where $\bx \in \RR^n$ is an input vector with dimension $n$,
$\bm{y} \in \RR^m$ is the output vector with dimension $m$, $\bm{W}
\in \RR^{m \times n}$ is a matrix of weights, and $\bm{b} \in \RR^m$
is a bias vector. 
Both $\bm{W}$ and $\bm{b}$ define the trainable
parameters of the network. The activation function $\pi$ is usually non-linear
(e.g., a rectified linear unit (ReLU)). 

A DNN $\mathbb{M}: R^n \mapsto R^m$ with $i$ hidden layers $\bm{h}$ can be formulated as:
\begin{align}
 \bm{h}_1 &= \pi(\bm{W}_1 \bm{x} + \bm{b}_1), &\nonumber\\
 \bm{h}_j &= \pi(\bm{W}_j \bm{h}_{j - 1} + \bm{b}_j), &\forall j \in \{2, 3, ..., i\} \nonumber  \\
 \bm{y}   &= \pi(\bm{W}_{i+1} \bm{h}_i + \bm{b}_{i+1})  & \label{generic_NN}
\end{align}
where $\bm{x} \in \RR^n$ and $\bm{y} \in \RR^m$ are the input and output vectors.
Learning DNN model $\mathbb{M}$ on a data set $T$ 
consists of finding the matrices $\bm{W_j}$
and bias vectors $\bm{b_j}$ for all $j \in \{1, 2, ... i+1\}$
to make the output predictions $\bm{\widehat{y_t}}$ close to the
ground truth data $\bm{y_t}$ for all $t \in T$, as
measured by a loss function $\mathbb{L}$:
\begin{align}
 \min_{\bm{W_j},\bm{b_j}: j \in [1, i+1]} \sum_{t \in T} \mathbb{L}(\bm{y_t}, \bm{\widehat{y_t}}), \nonumber \\
 \mbox{where } \bm{\widehat{y_t}} = \mathbb{M}(\bm{x_t})
\end{align}

\section{Regionally Decentralized AC-OPFs}
\label{sec:distributedOPFs}
This section presents the ADMM mechanism to solve 
Regionally Decentralized AC-OPFs. The presentation is largely based on~\cite{sun21two} and describes the regional AC-OPF model 
for each region, followed by showing the Augmented Lagrangian 
formulation for the ADMM approach.

\begin{model}[!t]
	{\small
	\caption{Regional AC Optimal Power Flow: $P_{RAC}$}
	\label{model:reg_ac_opf}
	\vspace{-6pt}
	\begin{align}
        \mbox{\bf input: } \;\; & \bm{S^d}[k] = ( S^d_i: i\in N_k) \nonumber \\
        & \bm{S^C}[k] = (S^C_{ij} : (i,j) \in R_k \cup R_k^R) \nonumber \\
        & \bm{V^C}[k] = (V^C_{i} : (i,j) \in R_k \cup R_k^R) \nonumber \\
		\mbox{\bf variables: } \;\;
		& \bm{S^g}[k] =(S^g_i: \forall i \in N_k),  \nonumber \\
		& \bm{V}[k] = (V_i: \forall i \in N_k \cup N^N_k) \nonumber \\
		& {\bm{S^f}[k] = (S^f_{ij} :  \forall(i,j)\in E_k \cup E_k^R \cup R_k \cup R_k^R )} \nonumber \\  		  
		\mbox{\bf minimize: } \;\;
		& {\cO}(\bm{S^g}[k]) = \sum_{i \in N_k} M_{i}(\Re(S^g_i)) \label{reg_ac_obj} 
	\end{align}	
	\begin{align}
		\mbox{\bf subject } & \mbox{\bf to intra-regional constraints:} \nonumber\\
		& \underline{v}_i \leq v_i \leq \overline{v}_i  		& \forall i \in N_k \label{eq:reg_ac_1} \\
		& \overline{S}^g_i \leq S^g_i \leq \underline{S}^g_i & \forall i \in N_k \label{eq:reg_ac_2}  \\
		& |S^f_{ij}| \leq \overline{S}_{ij} 					& \forall (i,j) \in E_k \cup E_k^R \label{eq:reg_ac_3}  \\
		& S^f_{ij} = {Y}^*_{ij} |V_i|^2 - {Y}^*_{ij} V_i V^*_j 			 & \forall (i,j)\in E_k \cup E^R_k \label{eq:reg_ac_4} \\
		& S^g_i - {S}^d_i = \displaystyle\sum_{(i,j)\in E_k \cup E_k^R} S^f_{ij} & \forall i\in N_k \setminus N^B_k\label{eq:reg_ac_5}  \\
		\mbox{\bf subject } & \mbox{\bf to inter-regional constraints:} \nonumber\\
		& |S^f_{ij}| \leq \overline{S}_{ij} 					& \forall (i,j) \in R_k \cup R_k^R \label{eq:reg_ac_6}  \\
		& S^f_{ij} = {Y}^*_{ij} |V_i|^2 - {Y}^*_{ij} V_i V^*_j 			 & \forall (i,j)\in R_k \cup R^R_k \label{eq:reg_ac_7} \\
		& S^g_i - {S}^d_i = \mkern-45mu \displaystyle\sum_{(i,j)\in E_k \cup E_k^R \cup R_k \cup R_k^R} S^f_{ij} & \forall i\in N^B_k\label{eq:reg_ac_8} 	\\
		\mbox{\bf subject } & \mbox{\bf to consensus constraints:} \nonumber \\
		&S^f_{ij} = S_{ij}^C & \forall (i,j) \in R_k \cup R_k^R \label{eq:reg_ac_9}\\
		&V_{i} = V_{i}^C  & \forall (i,j) \in R_k \cup R_k^R \label{eq:reg_ac_10}
	\end{align}
	}
	\vspace{-10pt}
\end{model}

\begin{figure}[!t]
\centering
\includegraphics[width=0.80\linewidth]{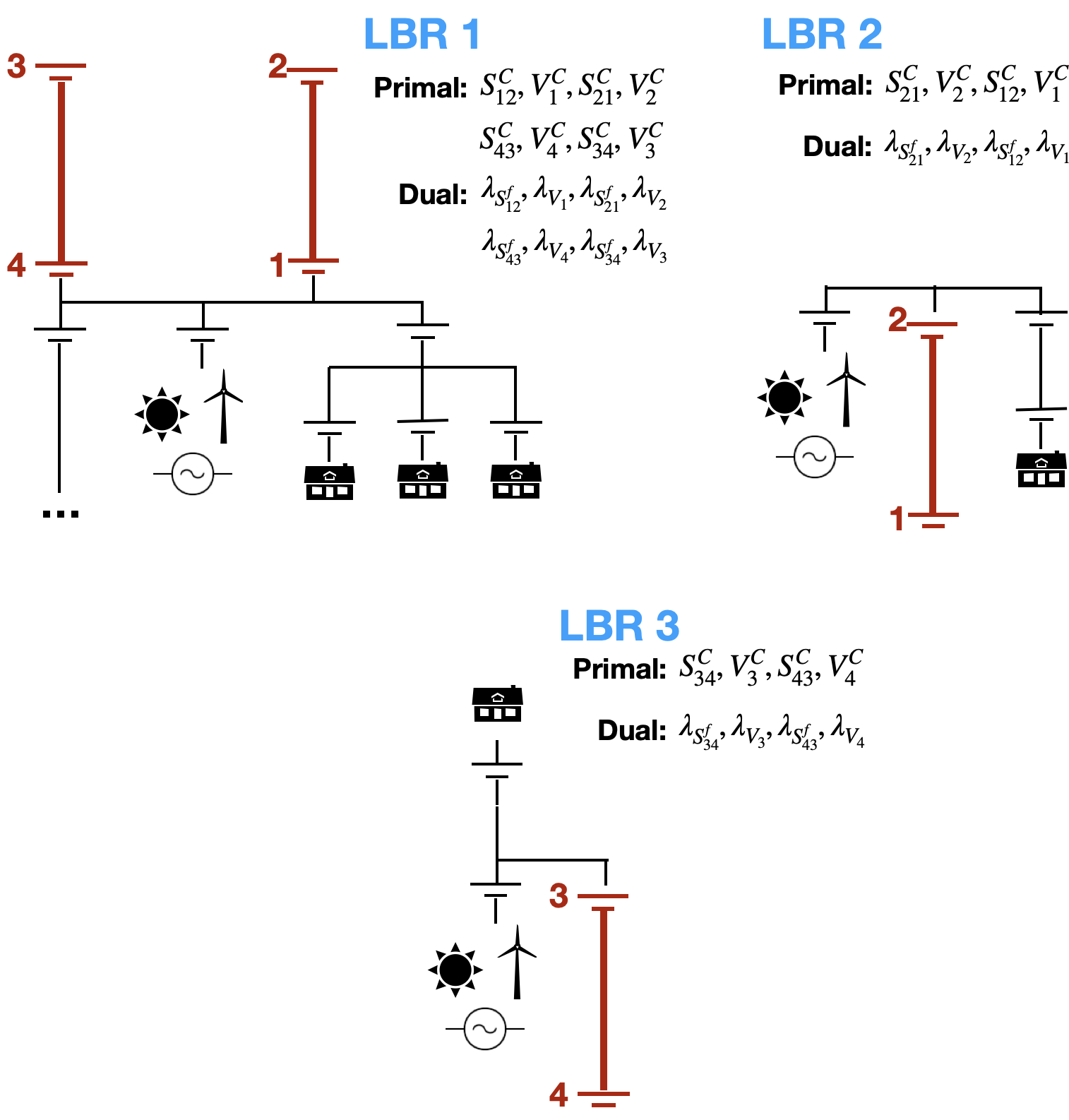}
\caption{An Example of Regional Decomposition.}
\label{fig:example2}
\end{figure}

\subsection{Regional Decentralized AC-OPF Model with Consensus}
Model~\ref{model:reg_ac_opf} 
presents the regional decentralized AC-OPF model  
for each load balancing region/zone $k \in K$, based on the centralized Model~\ref{model:ac_opf}. 
Figure~\ref{fig:example2} shows the decomposition diagram,
based on the example in Figure~\ref{fig:example1}.
Each load balancing region only considers 
the grid within their boundary, plus the interconnections (coupling branches and their associated buses).
The model relies on matching the consensus parameters 
$\bm{S^C}[k]$ and $\bm{V^C}[k]$ on the interconnections, by 
constraints \eqref{eq:reg_ac_9}-\eqref{eq:reg_ac_10},
to synchronize with the other regions.

\subsection{Augmented Lagrangian Reformulation}

Model~\ref{model:admm_ac_opf} shows the Augmented Lagrangian
relaxation for Model~\ref{model:reg_ac_opf}, with the introduction of
Lagrangian duals $\bm{\lambda_S}[k]$, $\bm{\lambda_V}[k]$, and the
$\rho$ penalty parameters for each load balancing region $k$.
Similarly, Model~\ref{model:two-level_admm_ac_opf} in the appendix shows 
the Augmented Lagrangian relaxation for the two-level ADMM 
(used by the inner loop).
These models will be used in the ADMM algorithm,
described in the next subsection.

\subsection{ADMM Algorithm}
The ADMM algorithm (Algorithm~\ref{alg:admm})
receives the network topology $\bm{\mathcal N}$, the load information $\bf{S}^d$, 
and the search parameters $\rho_0$ and $t_{max}$. 
Lines \ref{alg:init_start} - \ref{alg:init_end}
initialize the penalty parameter,
and initialize the
consensus variables and their corresponding Lagrangian duals for all the regions $k \in K$.
Line~\ref{alg:admm_iter_loop} executes the core procedure
$t_{max}$ times, and 
Line~\ref{alg:region_loop} iterates over each region $k$. 
Line~\ref{alg:region_opt} executes Model~\ref{model:admm_ac_opf}
for each region.
Line~\ref{alg:line_Lag_update} - \ref{alg:volt_Lag_update}
update the Lagrangian multipliers for each of the region. 
Finally, line~\ref{alg:line_con_update} - \ref{alg:volt_con_update}
export the consensus parameters.
The traditional (aka flat-start) ADMM procedure can be initialized as shown in Algorithm~\ref{alg:admm_init_cold}. 
The algorithm for the two-level ADMM is presented in the appendix (Algorithm~\ref{alg:two-level-admm}).

\begin{model}[h]
	{\small
	\caption{Augmented Lagrangian Regional AC-OPF: $P_{L}$}
	\label{model:admm_ac_opf}
	\vspace{-6pt}
	\begin{align}
        \mbox{\bf input: } \;\; & \bm{S^d}[k] = ( S^d_i: i\in N_k) \nonumber \\
        & \bm{S^C}[k] = (S^C_{ij} : (i,j) \in R_k \cup R_k^R) \nonumber \\
        & \bm{\lambda_S}[k] = (\lambda_{S^f_{ij}} : (i,j) \in R_k \cup R_k^R) \nonumber \\        
        & \bm{V^C}[k] = (V^C_{i} : (i,j) \in R_k \cup R_k^R) \nonumber \\
        & \bm{\lambda_V}[k] = (\lambda_{V_{i}} : (i,j) \in R_k \cup R_k^R) \nonumber \\                
        & \rho \nonumber \\  	
		\mbox{\bf variables: } \;\;
		& \bm{S^g}[k] =(S^g_i: \forall i \in N_k),  \nonumber \\
		& \bm{V}[k] = (V_i: \forall i \in N_k \cup N^N_k) \nonumber \\
		& {\bm{S^f}[k] = (S^f_{ij} :  \forall(i,j)\in E_k \cup E_k^R \cup R_k \cup R_k^R )} \nonumber \\  	
		\mbox{\bf minimize: } \;\;
		& \sum_{i \in N_k} M_{i}(\Re(S^g_i)) +\nonumber \\
		&  \mkern-25mu \displaystyle\sum_{(i,j)\in R_k \cup R_k^R} (\lambda_{S^f_{ij}} \cdot S^f_{ij}) + 
		   \mkern-20mu \displaystyle\sum_{(i,j)\in R_k \cup R_k^R} (\lambda_{V_{i}} \cdot V_{i}) + \label{eq:Lag_obj} \\
		&  \frac{\rho}{2} [\mkern-20mu \displaystyle\sum_{(i,j)\in R_k \cup R_k^R} \lVert S^f_{ij} - S_{ij}^C \rVert_2^2 + 
		   \mkern-25mu \displaystyle\sum_{(i,j)\in R_k \cup R_k^R} \lVert V_{i} - V_{i}^C \rVert_2^2 \;]  \nonumber\\		   
		\mbox{\bf subject to : }& \eqref{eq:reg_ac_1} - \eqref{eq:reg_ac_8} \nonumber
	\end{align}
	}
	\vspace{-10pt}
\end{model}

\begin{algorithm}[h]
\SetKwInOut{Input}{Inputs}
\SetKwInOut{Output}{Output}
\caption{ADMM: Main routine}
\label{alg:admm}
{\footnotesize

\SetKwInOut{InputN}{Network data}
\SetKwInOut{InputA}{Search parameters}
\SetKwInOut{InputP}{Primal initial input}
\SetKwInOut{InputD}{Dual initial input}
\SetKwInOut{Output}{Output}
\SetKwFunction{PL}{$P_{L}$}
\SetKwProg{Fn}{Function}{:}{}
\SetKwProg{Mn}{Algorithm}{:}{}

\InputN{$\bm{\mathcal N}, \bf{S}^d$}
\InputA{$\rho_0, t_{max}$}
$\rho \gets \rho_0$ \label{alg:init_start}\\
\For{$k \in K$}{
$\bm{S^C}[k], \bm{\lambda_S}[k], \bm{V^C}[k], \bm{\lambda_V}[k] \gets 
\texttt{initialize}(k)
$ \label{alg:init_end}\\
}

\For{$t = 1, 2, \ldots, t_{max}$} 
{ \label{alg:admm_iter_loop}
 \For{$k \in K$}{ \label{alg:region_loop}
   Regional AC-OPF:\\
   $(S^f_{ij}, V_i: (i,j) \in R_k \cup R_k^R) \gets $  \PL{$\bm{S^d}[k], \bm{S^C}[k], \bm{\lambda_S}[k], \bm{V^C}[k], \bm{\lambda_V}[k]$} \label{alg:region_opt}\\
   Lagrange multiplier update:\\
   $\bm{\lambda_S}[k] \gets ( \lambda_{S^f_{ij}} \gets \lambda_{S^f_{ij}} + (S^f_{ij} - S^C_{ij}) : (i,j) \in R_k \cup R_k^R)$ \label{alg:line_Lag_update}\\
   $\bm{\lambda_V}[k] \gets ( \lambda_{V_{i}} \gets \lambda_{V_{i}} + (V_{i} - V^C_{i}) : (i,j) \in R_k \cup R_k^R)$ \label{alg:volt_Lag_update}\\
   Consensus update:\\
   $S^C_{ij}  \gets (S^C_{ij} + S^f_{ij}) / 2 : \forall (i,j) \in R_k \cup R_k^R$  \label{alg:line_con_update} \\
   $V^C_{i}  \gets (V^C_{i} + V_i) / 2    : \forall (i,j) \in R_k \cup R_k^R$  \label{alg:volt_con_update}\\
 }
Penalty $\rho$ update (optional)\\
$\rho \gets $ update\_$\rho$()\\
}
}
\end{algorithm}
\begin{algorithm}[h]
\setcounter{AlgoLine}{0}
\SetKwInOut{Input}{Inputs}
\SetKwInOut{Output}{Output}
\caption{Cold-start  Initialization}
\label{alg:admm_init_cold}
{\footnotesize

\SetKwInOut{Output}{Output}
\SetKwProg{Fn}{Function}{:}{}
\SetKwProg{Init}{Initialize}{:}{}
\Fn{\texttt{initialize}(k)}{
  $\bm{S^C}[k] \gets (S^C_{ij} \gets 0 : (i,j) \in R_k \cup R_k^R)$\\ 
  $\bm{V^C}[k] \gets (V^C_{i} \gets 1 : (i,j) \in R_k \cup R_k^R)$\\
     $\bm{\lambda_S}[k] \gets ( \lambda_{S^f_{ij}} \gets 0 : (i,j) \in R_k \cup R_k^R)$ \\
     $\bm{\lambda_V}[k] \gets ( \lambda_{V_{i}} \gets 0 : (i,j) \in R_k \cup R_k^R)$\\
}
}
\end{algorithm}
\begin{algorithm}[h]
\setcounter{AlgoLine}{0}
\SetKwInOut{Input}{Inputs}
\SetKwInOut{Output}{Output}
\caption{Warm-start with ML}
\label{alg:admm_init_warm}
{\footnotesize

\SetKwInOut{Output}{Output}
\SetKwProg{Fn}{Function}{:}{}
\SetKwProg{Init}{Initialize}{:}{}
\Fn{\texttt{initialize}(k)}{
  $\bm{S^C}[k] \gets \hat{\bm{S}}^C[k] = (S^C_{ij} \gets \hat{S}^C_{ij} : (i,j) \in R_k \cup R_k^R)$\\ 
  $\bm{V^C}[k] \gets \hat{\bm{V}}^C[k] = (V^C_{i} \gets \hat{V}^C_{i} : (i,j) \in R_k \cup R_k^R)$\\
  $\bm{\lambda_S}[k] \gets \hat{\bm{\lambda}}_S[k] = ( \lambda_{S^f_{ij}} \gets \hat{\lambda}_{S^f_{ij}} : (i,j) \in R_k \cup R_k^R)$ \\
     $\bm{\lambda_V}[k] \gets \hat{\bm{\lambda}}_V[k] = ( \lambda_{V_{i}} \gets \hat{\lambda}_{V_{i}} : (i,j) \in R_k \cup R_k^R)$\\
}
}
\end{algorithm}

\section{Learning Architecture: ML-ADMM}
\label{sec:distributedlearning}

The previous section presented how to utilize decentralized
optimization, e.g., ADMM methods in Algorithm~\ref{alg:admm}, to find flows and voltages for
shared interconnections. 
This section presents a decentralized
machine-learning approach to speed up ADMM search by learning these entities. 

\subsection{Overview}
The machine-learning approach is motivated by the recognition that, in
practice, it would be costly to cold-start the ADMM instead of using
predictions for the consensus variables ($\bm{S}^C$,
$\bm{V}^C$) and their corresponding dual multipliers
($\bm{\lambda}_{S}$, $\bm{\lambda}_{V}$). 
If these predictions are available, the ADMM procedure can be initialized as in Algorithm~\ref{alg:admm_init_warm}. If all the consensus variables and dual multipliers are perfectly predicted, only one ADMM iteration would be required.

Predictions on load demands
and renewable generations are already incorporated by various ISOs in
their markets (e.g., MISO~\cite{MISO_BPM}). {\em ML-ADMM generalizes
  this practice by incorporating the predictions on the consensus
  variables.}
The proposed methodology was applied within a general ADMM framework and the nonlinear AC-OPF formulation to demonstrate how to develop learning strategies for learning decentralized
optimization problems in power systems. Note that the
proposed learning methodology is general and does not necessarily
require an augmented Lagrangian formulation and/or an ADMM
approach. 
The same approach can be applied on other types of regional decomposition algorithms, with other types of OPF formulations, e.g., DC/linearized formulation or second-order cone OPF formulation. 
In addition, this approach does not change the inherent  computational complexity of the underlying decomposition framework, nor does it modify any of existing communication architectures. The only addition is that agents need to train their own learning framework to initialize the underlying decomposition.

The remaining subsections will introduce the machine learning architecture (ML-ADMM) to predict the necessary quantities for Algorithm~\ref{alg:admm_init_warm}, and how to train the machine learning models.

\subsection{Deep Learning Models}
ML-ADMM aims at learning two sets of parameters: the consensus variables $\bm{S}^C$, $\bm{V}^C$ and their corresponding dual multipliers $\bm{\lambda}_{S}$, $\bm{\lambda}_{V}$ for every region $k$,
based on the current load forecast $\bm{S^d}$.
Since these parameters are complex quantities, ML-ADMM first 
splits each quantity into its individual components as follows:
\begin{align}
    &\bm{S^d} \mapsto \bm{p^d} + i \bm{q^d}, \nonumber\\
    &\bm{S}^C \mapsto \bm{p}^C + i \bm{q}^C, \;\; \bm{V}^C \mapsto \bm{v}^C \angle \bm{\theta}^C, \nonumber \\
    &\bm{\lambda_S} \mapsto \bm{\lambda_p} + i \bm{\lambda_q}, \;\; \bm{\lambda_V} \mapsto \bm{\lambda_v} \angle \bm{\lambda_\theta}, \nonumber
\end{align}
\noindent
where $\bm{X} \mapsto \bm{X}_r + i \bm{X}_i$ splits a complex vector $\bm{X}$ into the real component vector $\bm{X}_r$ and the imaginary component vector $\bm{X}_i$ (i.e., splitting into the rectangular form), and 
$\bm{X} \mapsto \bm{X}_m + \angle \bm{X}_\theta$
splits the complex vector $\bm{X}$ into the magnitude vector $\bm{X}_m$ and the angle vector $\bm{X}_\theta$ (i.e., splitting into the polar form).
Let $\bm{x}$ be the flattened input vector ($\bm{p^d}, \bm{q^d}$), and
$\bm{y}[k]$ to be the target prediction quantities, where
\begin{align}
    \bm{y}[k] & = 
    \begin{cases}
        \bm{p}^C[k], \mbox{ for active line flow} \\
        \bm{q}^C[k], \mbox{ for reactive line flow} \\
        \bm{v}^C[k], \mbox{ for voltage magnitude} \\
        \bm{\theta}^C[k], \mbox{ for voltage angle} \\
        \bm{\lambda_p}[k], \mbox{ for active line flow dual} \\
        \bm{\lambda_q}[k], \mbox{ for reactive line flow dual} \\
        \bm{\lambda_v}[k], \mbox{ for voltage magnitude dual, and} \\
        \bm{\lambda_\theta}[k], \mbox{ for voltage angle dual} \\
    \end{cases}\label{dnn_type}
\end{align}
for each load balancing zone/region $k$. 
To initialize the ADMM algorithm with predictions (Algorithm~\ref{alg:admm_init_warm}), 
each region $k$ will only need to 
learn and predict all 8 types of $\bm{y}[k]$ independently, based on the
current system load demand (input feature vector $\bm{S^d}$).

\begin{figure}[!t]
\centering
\includegraphics[width=0.80\linewidth]{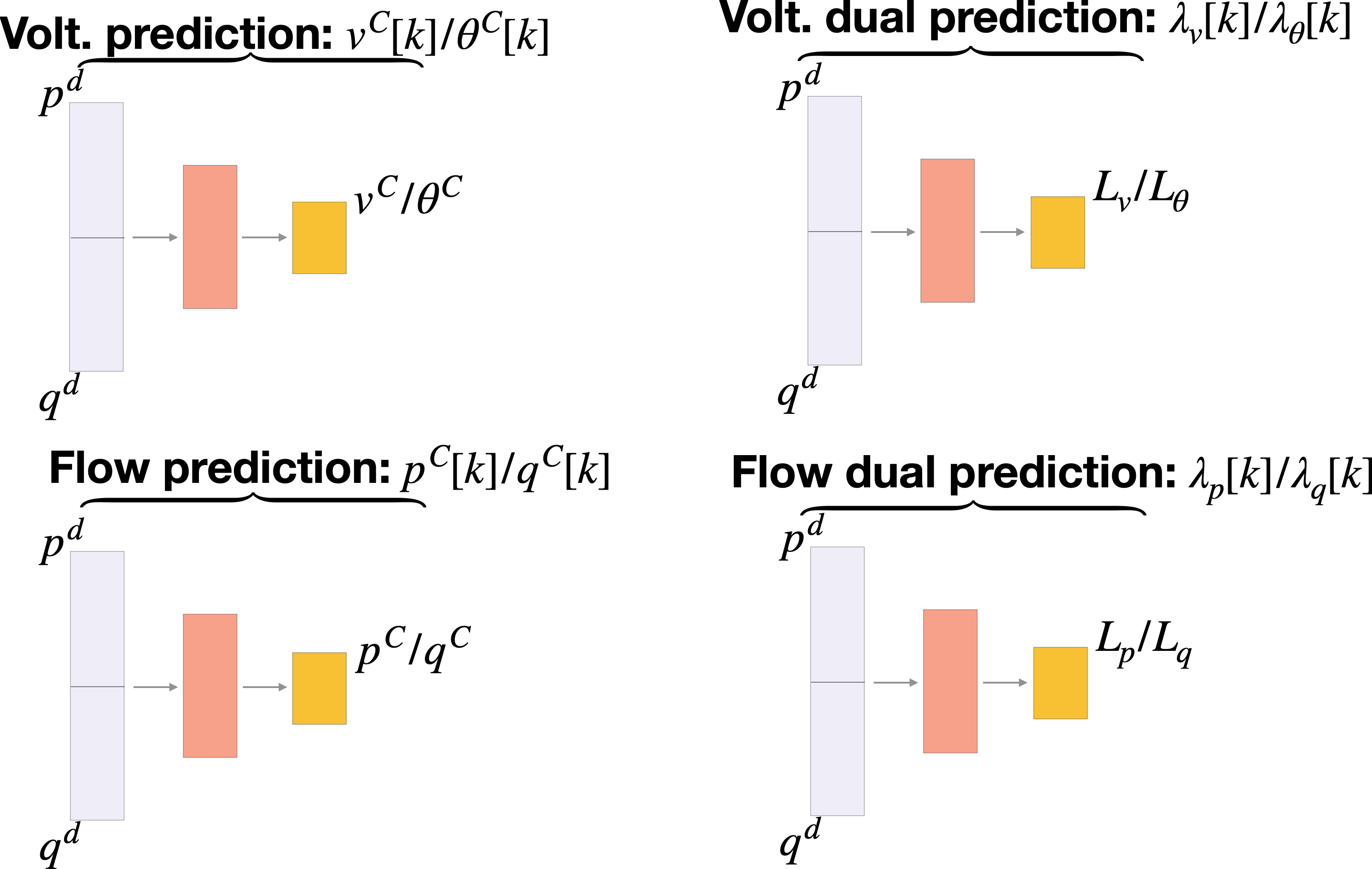}
\caption{DNNs for the coupling parameters and the associated dual multipliers. }
\label{fig:NN_generic}
\end{figure}

In order to achieve the task, ML-ADMM constructs DNNs $\mathbb{M}_{\bm{y}[k]}$
of the form:
\begin{align}
 \mathbb{M}_{y[k]}(\bm{x}): \bm{y[k]} = \pi(\bm{W}_2 \bm{h} + \bm{b}_2), \mbox{with } \bm{h}    = \pi(\bm{W}_1 \bm{x} + \bm{b}_1)  \nonumber
\end{align}
where $\bm{h}$ is the hidden layer with a dimension set to twice the dimension of the output vector $\bm{y}[k]$. Figure~\ref{fig:NN_generic} illustrates the four types of DNNs constructed by ML-ADMM for predicting the coupling parameters and the associated dual multipliers for each region $k$.

\subsection{Decentralized Training}
ML-ADMM trains the models $\mathbb{M}_{\bm{y}[k]}$ of each load
balancing zone/region $k$ in a decentralized fashion using a data set
$T[k]$, owned by region $k$. 
The training is performed in parallel for each type of target prediction quantity (listed in \eqref{dnn_type}).
Algorithm~\ref{alg:dnn_training} showcases
a high-level view of the training process with back-propagation.
The training terminates when $e_{max}$ epochs
have been executed (Line~\ref{alg:train:start}).
For each epoch, the algorithm then iterates over every pair of input and output features $(\bm{x}, \bm{y}[k])$ from the data set $T[k]$  (Line~\ref{alg:train:data}). 
Line 3 computes the prediction for the target quantities.
Finally, line 4 updates the DNN using back-propagation (\texttt{BACKPROP}), based on the current prediction error measured by the loss function $\mathbb{L}$.
After all the models $\mathbb{M}_{\bm{y}[k]}$ are trained by ML-ADMM,
Algorithm~\ref{alg:admm_init_warm}
can then be applied with the
predicted quantities. 
{\em Observe that both the training and the
  optimization proceeds in a fully decentralized fashion. Moreover,
  during training, the region do not need to interact with each other.}

\begin{algorithm}[!t]
\setcounter{AlgoLine}{0}
\SetKwInOut{Input}{Inputs}
\SetKwInOut{Output}{Output}
\caption{Regional Training with Backpropagation}
\label{alg:dnn_training}
{\footnotesize

\SetKwInOut{Output}{Output}
\SetKwProg{Fn}{Function}{:}{}
\SetKwProg{Init}{Initialize}{:}{}
\Input{Initialized $\mathbb{M}_{\bm{y}[k]}$; Max epoch $e_{max}$; Data set $T[k]$}
\For{$e = 1, 2, \dotsc, e_{max}$}{\label{alg:train:start}
  \For{$(\bm{x}, \bm{y}[k]) \in T[k]$}{\label{alg:train:data}
  $\widehat{\bm{y}[k]} \gets \mathbb{M}_{\bm{y}[k]}(\bm{x})$ \label{alg:train:predict}\\
  $\mathbb{M}_{\bm{y}[k]} \gets \textsc{BackProp}(\mathbb{L}(\widehat{\bm{y}[k]}, \bm{y}[k]))$ \label{alg:train:end}
  }
}   
}
\end{algorithm}

\begin{figure*}[!t]
\centering
\includegraphics[width=0.23\linewidth]{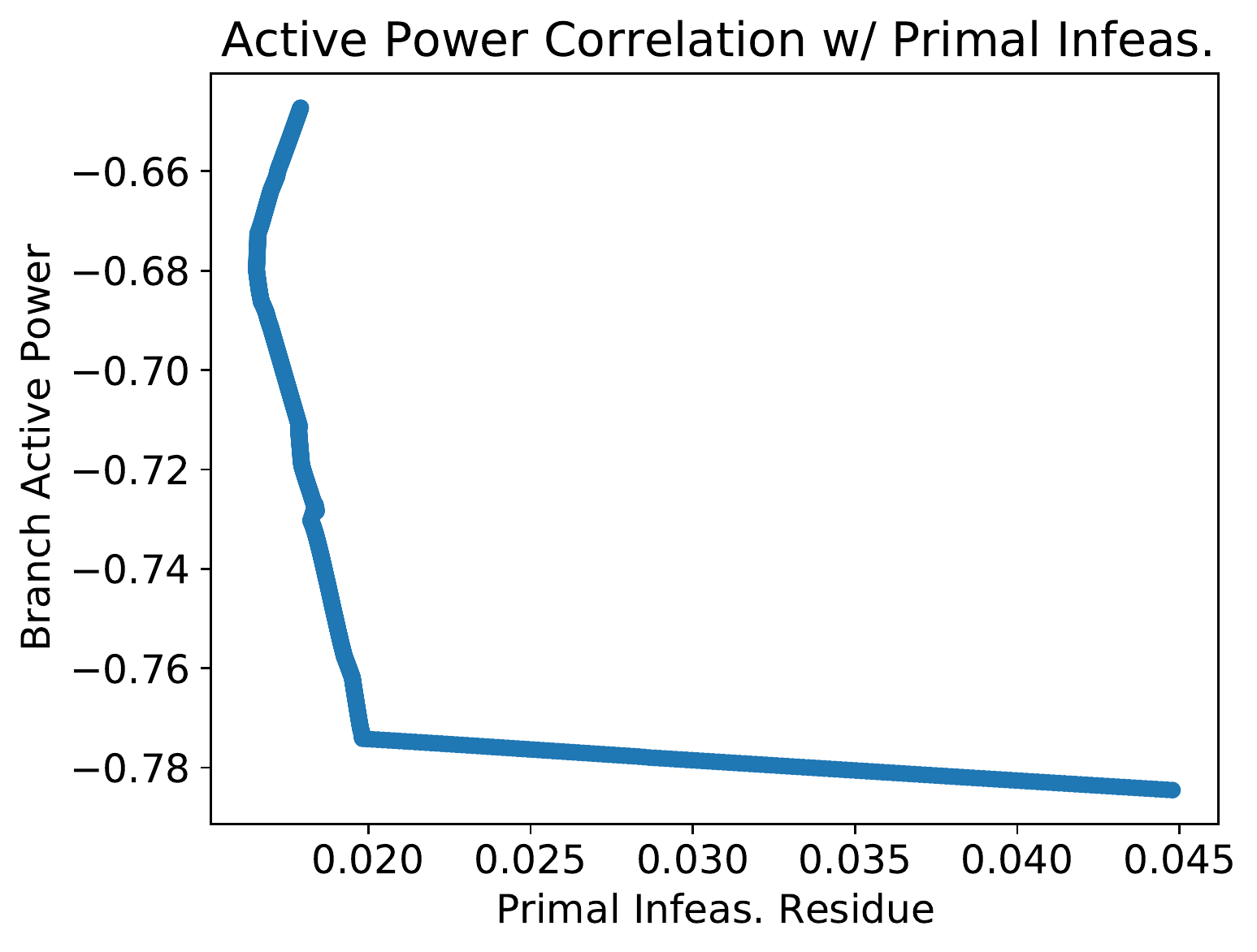}
\includegraphics[width=0.23\linewidth]{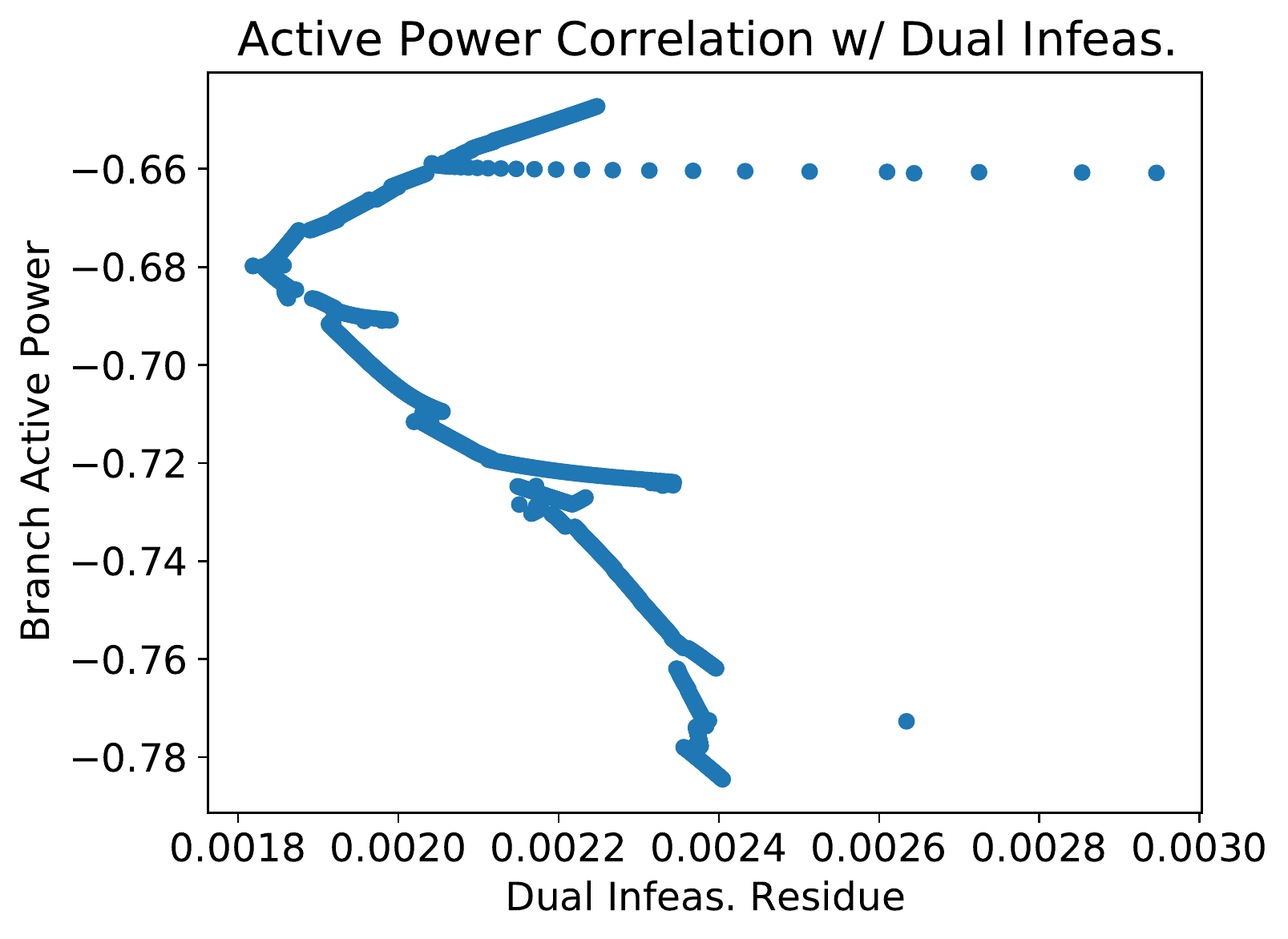}
\includegraphics[width=0.23\linewidth]{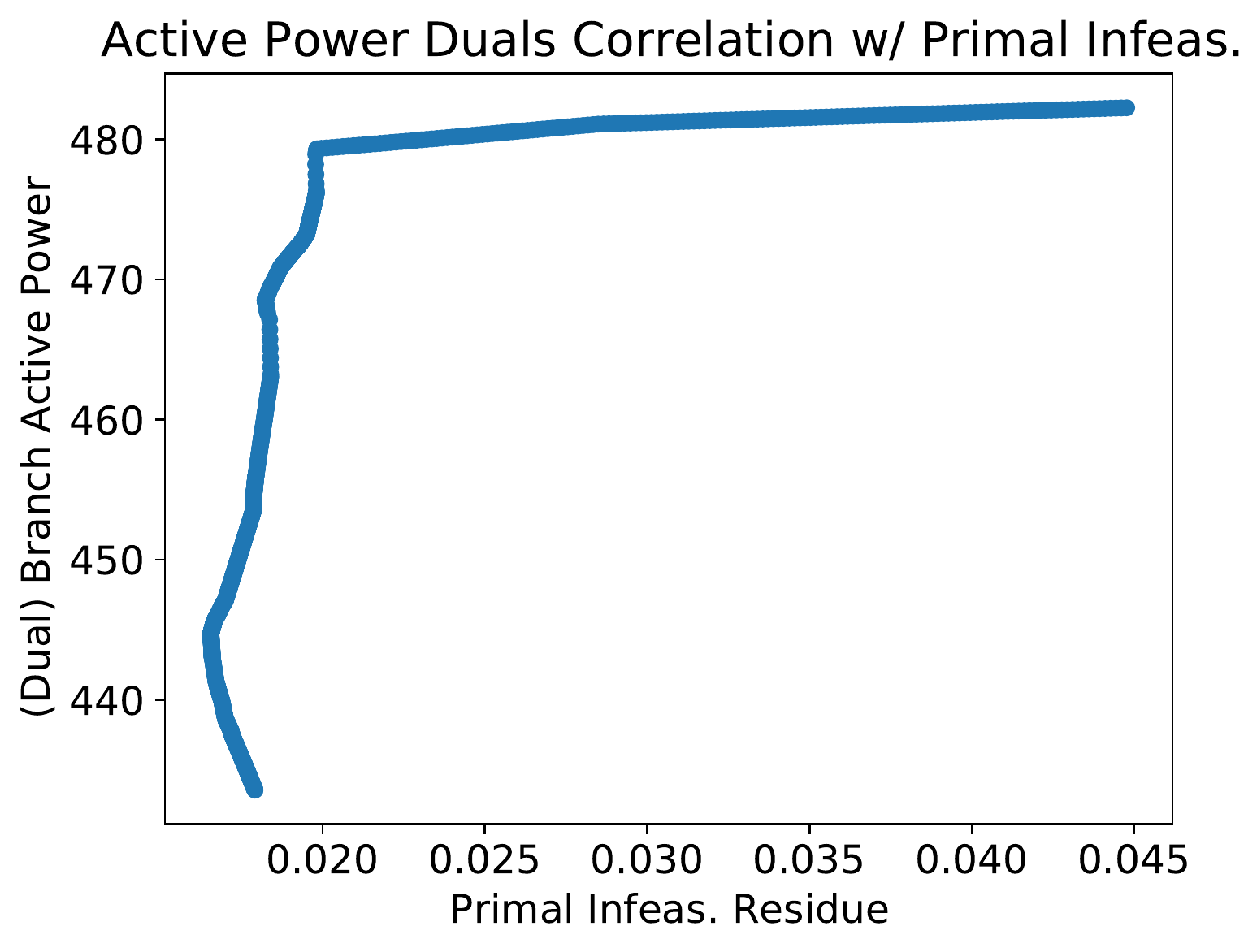}
\includegraphics[width=0.23\linewidth]{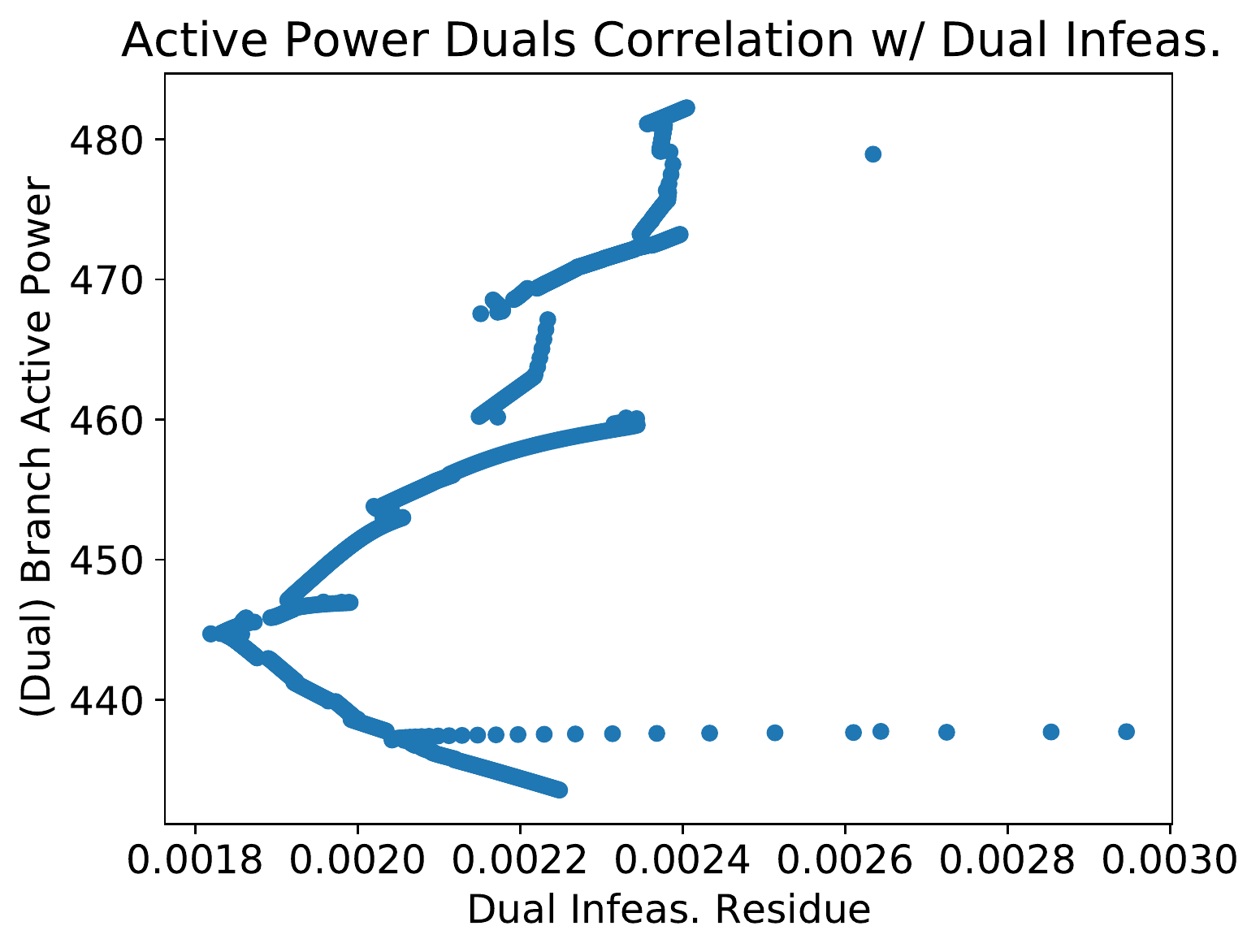}
\caption{Active power and its duals at a coupling branch, sorted by primal/dual infeasibility residues. (p.u.)}
\label{fig:infeas_ranking1}
\end{figure*}
\begin{figure*}[t]
\centering
\includegraphics[width=0.23\linewidth]{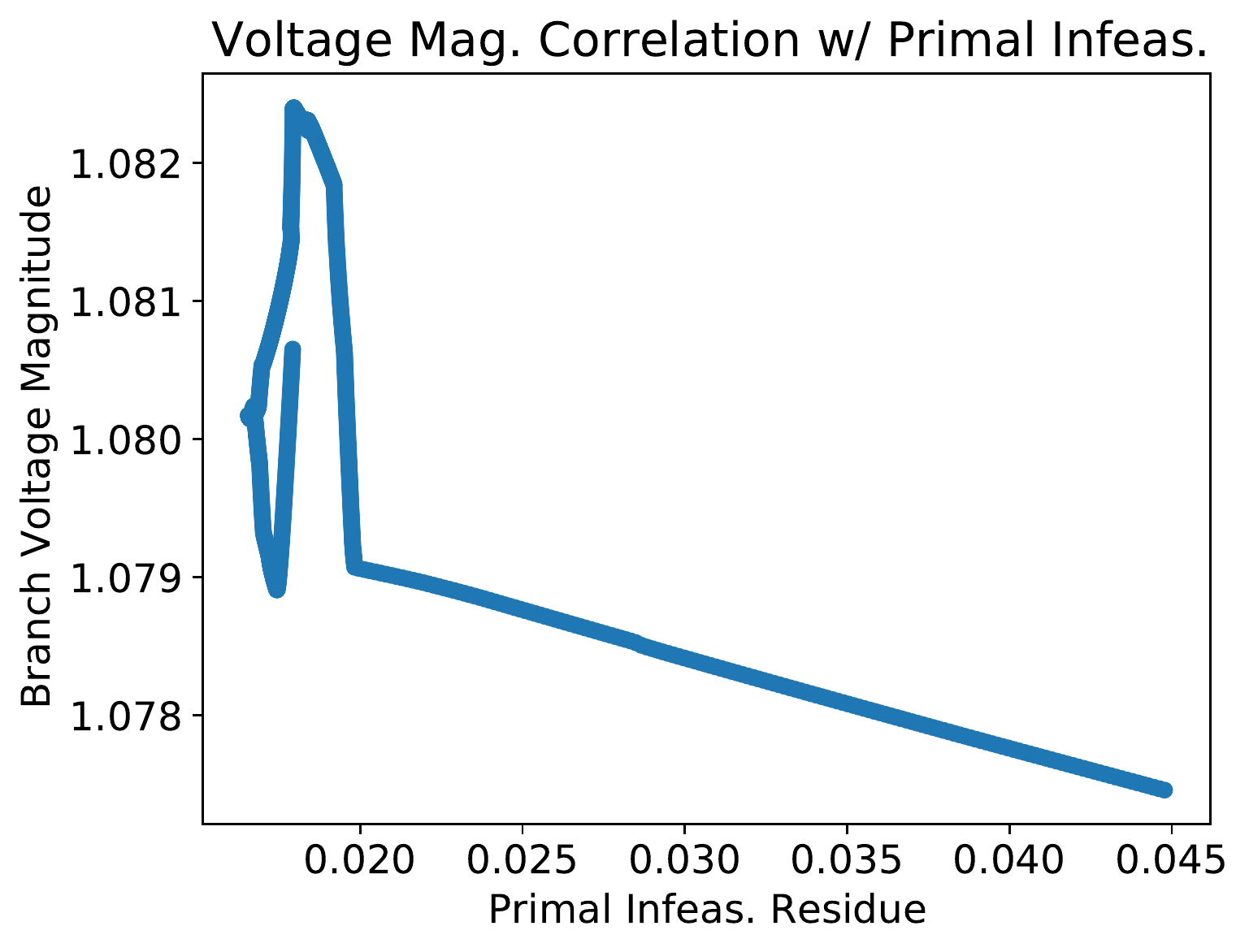}
\includegraphics[width=0.23\linewidth]{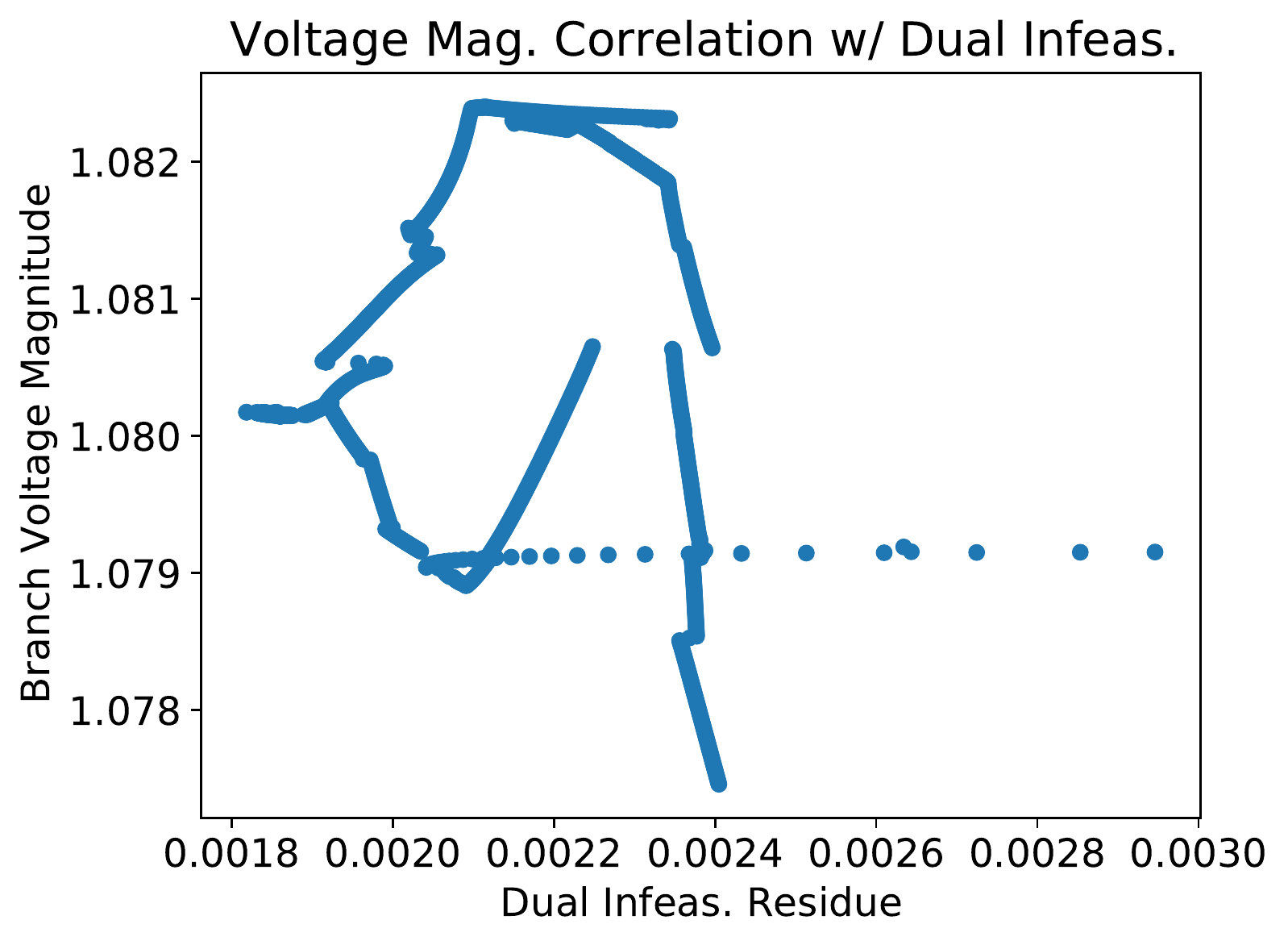}
\includegraphics[width=0.23\linewidth]{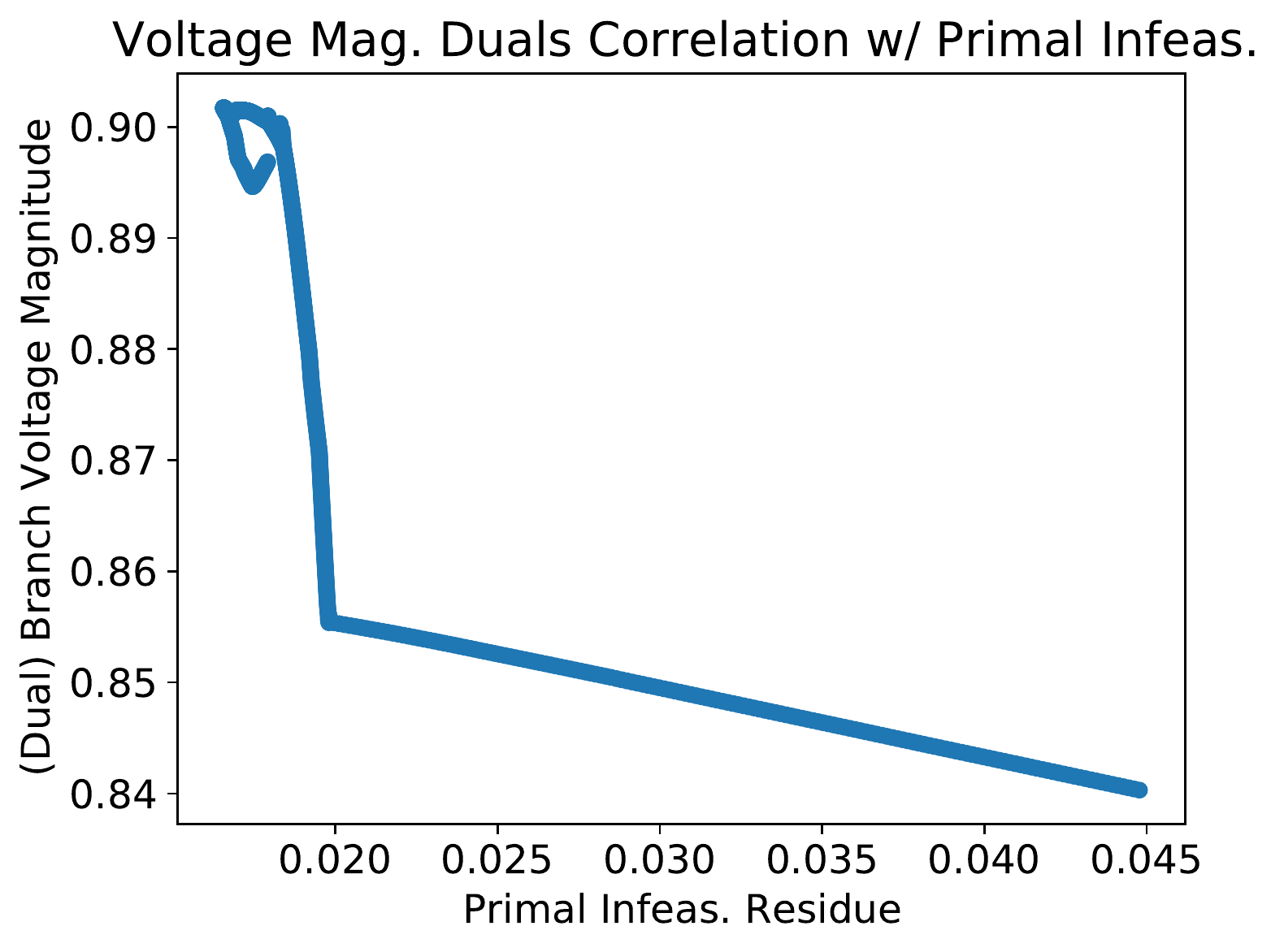}
\includegraphics[width=0.23\linewidth]{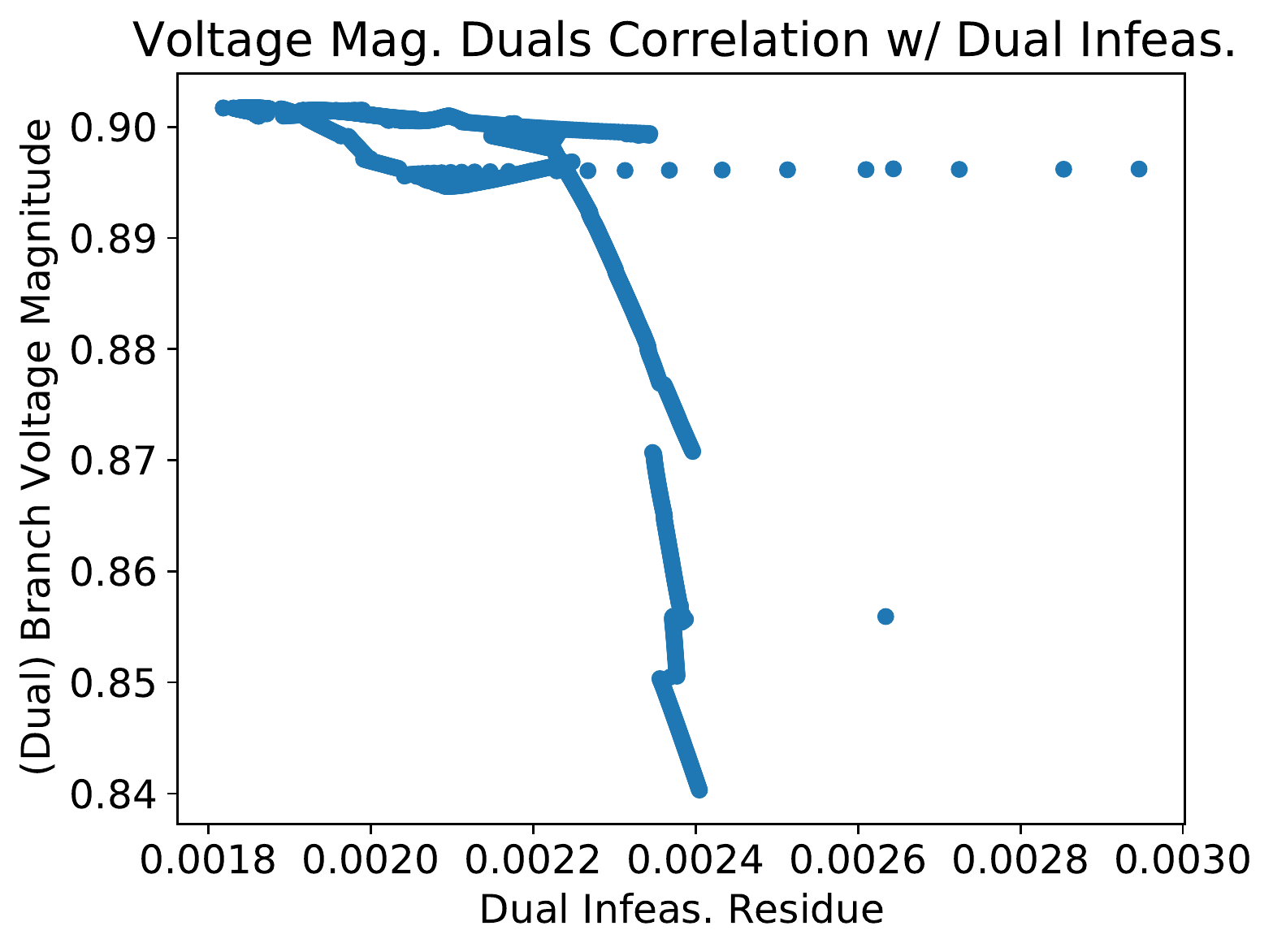}
\caption{Voltage magnitude and its duals at a coupling bus, sorted by primal/dual infeasibility residues. (p.u.)}
\label{fig:infeas_ranking2}
\end{figure*}

\subsection{Extension for Two-level ADMM}
As mentioned earlier, the two-level ADMM introduces slack variables
$z$ and their corresponding dual multipliers $\Lambda$ for all the
coupling constraints.  Let $\bm{z_p}$, $\bm{z_q}$, $\bm{z_v}$,
$\bm{z_\theta}$ to be the vector of slack variables for coupling
constraints corresponding to consensus variables $\bm{p}^C$,
$\bm{q}^C$, $\bm{v}^C$, and $\bm{\theta}^C$, and let $\bm{\Lambda_p}$,
$\bm{\Lambda_q}$, $\bm{\Lambda_v}$, $\bm{\Lambda_\theta}$ to be the
corresponding vector of dual multipliers for $\bm{z_p}$, $\bm{z_q}$,
$\bm{z_v}$, and $\bm{z_\theta}$.  Let $\bm{x}$ be the flattened input
vector ($\bm{p^d}, \bm{q^d}$), and $\bm{y}[k]$ to be the target
prediction quantities: $\bm{z_p}[k]$, $\bm{z_q}[k]$, $\bm{z_v}[k]$,
$\bm{z_\theta}[k], \bm{\Lambda_p}[k]$, $\bm{\Lambda_q}[k]$,
$\bm{\Lambda_v}[k]$, $\bm{\Lambda_\theta}[k]$, for each load balancing
zone/region $k$. The generalization of ML-ADMM learns and predicts the
8 additional types of quantities, based on the current system load
demand.  DNN architectures similar to those in
Figure~\ref{fig:NN_generic} are built and trained in the same
decentralized fashion.

\section{Filtering the Training Data}
\label{sec:data_filtering}
ADMM runs may exhibit significant convergence properties, even for closely related inputs. {\em The section investigates a novel idea: only using historical ADMM runs with high-quality convergence properties in the training set.} The motivation here is not to imitate the behavior of all ADMM runs: rather it is to find initial values for the consensus parameters that will enable strong convergence of the optimization model.

\paragraph{ADMM Behavior} The filtering idea is motivated
by the fact that ADMM runs for specific inputs may not be optimal,
unique, or may not have converged when reaching their termination
condition. For instance, Figures~\ref{fig:infeas_ranking1}
and~\ref{fig:infeas_ranking2} display the active power and voltage
magnitude, together with their corresponding dual multipliers, for a
specific coupling branch and one of its associated buses. The results
are for ADMM runs with 3,000 iterations over a variety of instances of
the France\_EHV test case. Each dot in the figures is an instance and
the figures report correlations between the primal \& dual
infeasibility residues on the one hand and the active power and
voltage magnitude (and their duals) on the other hand. As can be
observed, there are strong correlations between these quantities and
the convergence mesures (i.e., primal and dual infeasibility residues)
and also natural breakpoints that separates the runs with good
convergence properties from the more problematic runs. Learning from
instances with poor convergence qualities is not desirable and hence
ML-ADMM utilizes two set of filters to select the training data.

\paragraph{Convergence Filter}

The convergence filter $c(\alpha)$ returns a subset of the data set
$T[k]$ by filtering instances whose primal or dual infeasibility
residues are higher than a threshold specified by $\alpha$. In other
words, the convergence filter excludes data sets by splitting the
x-axis of Figures~\ref{fig:infeas_ranking1}
and~\ref{fig:infeas_ranking2}. Let $r_p(t)$ and $r_d(t)$ to be the
primal and dual infeasibility residue for instance $t$, and $A_{r_p}$
and $A_{r_d}$ to be two arrays storing, in ascending order, the primal
and dual infeasibility residues for all instances in $T[k]$. Let
$A_{r_p}[i]/A_{r_d}[i]$ to be the $i^{\mbox{th}}$ element of the array
$A_{r_p}/A_{r_d}$, and
\texttt{ceil} to be the ceiling function. %
The threshold
$r_p^{\mbox{thres}}$/$r_d^{\mbox{thres}}$ for primal and dual
infeasibility residues are given by:
\begin{align}
    r_p^{\mbox{thres}} = A_{r_p}[ \;\;\texttt{ceil}( \alpha \times \lvert T \rvert ) \;\; ] \nonumber \\
    r_d^{\mbox{thres}} = A_{r_d}[ \;\; \texttt{ceil}( \alpha \times \lvert T \rvert ) \;\; ] \nonumber
\end{align}
where $0 < \alpha \leq 1$. The data set returned by filter $c(\alpha)$ for region/zone $k \in K$ is thus
\begin{align}
    \{t \in T[k] \mbox{ where } r_p(t) \leq r_p^{\mbox{thres}} \wedge r_d(t) \leq r_d^{\mbox{thres}} \}. \nonumber
\end{align}
For the two-level ADMM, the convergence distance can be approximated by 
the magnitude of the slack parameters $\bm{z}$.
Therefore, the convergence filter $c(\alpha)$ for two-level ADMM 
can be simplified and returns a subset of the data set
$T[k]$ by filtering instances whose slack parameter distances, as measured by a metric scoring function, are higher than a threshold specified by $\alpha$.
Let $s_z(t)$ to be the
slack distance score for instance $t$, and $A_{s_z}$
to be the arrays storing, in ascending order, the scores
for all instances in $T[k]$. Let
$A_{s_z}[i]$ to be the $i^{\mbox{th}}$ element of the array
$A_{s_z}$, and
\texttt{ceil} to be the ceiling function. 
The threshold are given by:
\begin{align}
    s_z^{\mbox{thres}} = A_{s_z}[ \;\;\texttt{ceil}( \alpha \times \lvert T \rvert ) \;\; ] \nonumber 
\end{align}
where $0 < \alpha \leq 1$. The data set returned by filter $c(\alpha)$ for region/zone $k \in K$ for the two-level ADMM is thus
\begin{align}
    \{t \in T[k] \mbox{ where } s_z(t) \leq s_z^{\mbox{thres}} \}. \nonumber
\end{align}
Experimental evaluation uses the following slack distance scoring function $s_z(t)$:
\begin{align}
     \max(\lvert\bm{z_p}[t]\rvert) +
     \max(\lvert\bm{z_q}[t]\rvert) +
     \max(\lvert\bm{z_v}[t]\rvert) +
     \max(\lvert\bm{z_\theta}[t]\rvert). \nonumber     
\end{align}

\paragraph{Standard Deviation Filter}
The standard deviation filter $s(\beta)$ returns a subset of the data set $T[k]$ by filtering instances whose consensus/dual multiplier values are outliers. In other words, the filter excludes instances by splitting the y-axis of Figures~\ref{fig:infeas_ranking1} and~\ref{fig:infeas_ranking2}. Let $y[k](t)$ to be the target prediction quantity (either $\bm{p}^C[k]$, $\bm{q}^C[k]$, $\bm{v}^C[k]$, $\bm{\theta}^C[k]$,
$\bm{\lambda_p}[k]$, $\bm{\lambda_q}[k]$, $\bm{\lambda_v}[k]$, or $\bm{\lambda_\theta}[k]$) for instance $t \in T[k]$.
Let $m(y[k])$ and $\sigma(y[k])$ to be the mean and standard deviation vector of $y[k]$ across the data set $t \in T[k]$,
i.e., the mean and standard deviation for the set $\{y[k](t) : t \in T[k] \}$.
The data set returned by filter $s(\beta)$ for region/zone $k \in K$ is: 
\begin{align}
    \{t \in T[k] \mbox{ where } \lvert y[k](t) - m(y[k]) \rvert \preccurlyeq \beta \sigma(y[k])  \}, \nonumber
\end{align}
for all $y[k] \in \{$ $\bm{p}^C[k]$, $\bm{q}^C[k]$, $\bm{v}^C[k]$, $\bm{\theta}^C[k]$,
$\bm{\lambda_p}[k]$, $\bm{\lambda_q}[k]$, $\bm{\lambda_v}[k]$, $\bm{\lambda_\theta}[k]$ $\}$, where $\preccurlyeq$ generalizes $\leq$ for vectors.

\section{Experimental Evaluations}
\label{sec:evaluation}

This section presents the data-generation process, the implementation
and training details, the prediction accuracy, and convergence results
of the learning-boosted ADMM with respect to the original and
two-level ADMM and the AC-OPF solution.

\subsection{Experimental Setup}

\paragraph{Benchmarks}

The experiments were performed on three real benchmark
  networks: France\_EHV, LYON, and France.  All three of them were
  extracted from the French Transmission Grid, covering 12
  georgraphical regions of France. The experiments associate one agent
  per region.  
\footnote{
While the learning framework in principle 
should work on any number of regions/agents, 
artificially varying region/agent numbers on industrial systems 
may create unrealistic benchmarks/scenarios   
and breaking away from real-life practices 
(e.g., creating artificial inter-regional lines). 
}
These benchmarks contain 1700 to 6700 buses, and 140 to 320 coupling branches (regional interconnections). 
Table~\ref{table:networks} shows a summary of the benchmark statistics.
Detailed network parameters can be
found in~\cite{chatzos21spatial}.

\begin{table}[t]
\caption{Benchmark Networks}
    \begin{center}
    \resizebox{0.99\linewidth}{!}{
        \begin{tabular}{|c |c c c c|c c| c|} 
         \hline
        Benchmark & $|N|$ & $|E|$ & $|D|$ & $|G|$ & $|K|$ &  $|\bigcup_{k \in K} R_k|$ & Nom. Load\\ [0.5ex] 
        \hline\hline
        {\tt France\_EHV} & 1737 & 2350 & 1731 & 290 & 12 & 148 & 51949 MW\\ 
        \hline
        {\tt LYON} & 3411 & 4499 & 3273 & 771 & 12 & 219& 52394 MW\\
        \hline
        {\tt France} & 6705 & 8962 & 6262 & 1708 & 12 & 326&54708 MW\\
        \hline
        \end{tabular}}
\end{center}
\label{table:networks}
\end{table}

\paragraph{Implementation Details}

The ADMM and AC-OPF solving routines were implemented in Julia 1.6.1,
with Ipopt 3.12.13 (w/ HSL MA57) as the nonlinear solver. The learning
models were implemented in PyTorch \cite{paszke:17} and run with
Python 3.6, with the Mean Squared Error (MSE) as the loss
function. The training was performed in parallel on Intel CPU cores at
2.1GHz, one core for each region. The training used Averaged
Stochastic Gradient Descent (ASGD),with 64 mini-batches, 1000 epoches,
and 0.001 learning rate.

\paragraph{Data Generation}

The training data sets were generated by varying the load profiles of
each test network from 80\% to 122\% of their original (complex) load
values, with a step size of 0.01\%, giving a maximum of 4200 test
cases for every benchmark network.  For the two-level
  ADMM, the data sets were generated by varying the load profiles of
  each test network from 80\% to 120\% of their original load values
  with a step size of 0.02\%.  For each test case, to create enough
diversity, every load is perturbed with random noise from the polar
Laplace distribution whose parameter $\lambda$ is set to 1\% of the
apparent power. Test cases with no feasible AC solutions were removed
from the data set. The outputs of each test case is obtained by
running the implemented ADMM routine for 3000 iterations with $\rho$
set to 10 and 2000 iterations with $\rho = 1000$ for
 the two-level ADMM similar to~\cite{sun21two}.  Results from the ADMM
routine were recorded as the ground truth, and split with 80\%-20\%
ratio for training and testing purposes.

\paragraph{Evaluation Details}

The evaluation aims at determining whether machine learning can 
speed up the convergence of the ADMM. It compares the learning-boosted
ML-ADMM with three key baselines:
\begin{enumerate}
    \item Nominal initialization [N-ADMM] --- 
    the ADMM initialized with cold-start (nominal consensus and zero dual values) and
    run for 500 iterations;
    \item Ground Truth Data [P-DATA] --- 
    the ADMM initialized with cold-start (nominal consensus and zero dual values) and run for 3000 iterations. 
    \item Perfect initialization [P-ADMM] --- 
    the ADMM initialized with the perfect warm-start (ground-truth consensus and dual values) and 
    run for 500 iterations ($\approx$ P-DATA + ADMM).
\end{enumerate}
N-ADMM is used to assess whether ML-ADMM is effective in producing
solutions of better quality than nominal initializations with the same
small number of iterations. P-DATA is used to assess whether 500
iterations of ML-ADMM is effective in recovering solutions of the same
quality as ADMM with 3000 iterations (i.e., 1/6 of the original 3000
iterations). P-ADMM, which is seeded with the ground-truth data, is
used to measure how well ML-ADMM would perform in comparison with a
hot start with perfect information.

\begin{table}[h]
\caption{The Prediction Errors in Percentage for Various Filters. \label{tab:pred_error}}
\resizebox{0.99\linewidth}{!}
	{
	\begin{tabular}{| c | l | r r | r  r| r r| r r|}
    \toprule
    Network & Filter & $\bm{p}^C$ & $\bm{q}^C$ & $\bm{v}^C$ & $\bm{\theta}^C$ & 
                       $\lambda_{\bm{p}}$ & $\lambda_{\bm{q}}$ & $\lambda_{\bm{v}}$ & $\lambda_{\bm{\theta}}$ \\     
    \midrule
     \multirow{9}{*}{France\_EHV} & NIL & 8.25 & 16.17 & 5.23 & 12.71 & 2.36 & 10.26 & 18.14 & 11.32\\
     \cmidrule{2-10}
      & c(90\%) & 7.78 & 14.48 & 4.73 & 12.73 &2.16 & 8.64 & 15.98 & 10.49\\
                                 & c(80\%) & 7.56 & 13.26 & 4.34 & 12.65 & 1.99 & 7.86 & 14.71 & 9.86\\
                                 & c(70\%) & 7.17 & 12.03 & 3.95 & 12.20 & \bf{1.97} & 7.48 & 14.25 & 9.00 \\
                                 & c(60\%) & 6.44 & 10.46 & 3.41 & 10.78 & 2.07 & 7.12 & 13.40 & 7.71 \\
                                 & c(50\%) & \bf{5.77} & \bf{8.99} & \bf{2.89} & \bf{9.24} & 2.24 & \bf{6.66} & \bf{12.09} & \bf{6.54} \\
    \cmidrule{2-10}
                                 & s(4.0) & 8.14 & 16.00 & 5.14 & 12.56 & 2.34 & 10.04 & 17.58 & 11.13 \\
                                 & s(3.0) & 7.78 & 15.11 & 4.84 & 11.96 & 2.28 & 9.42 & 16.39 & 10.62 \\
                                 & s(2.0) & 6.89 & 13.05 & 4.08 & 10.82 & 2.01 & 7.52 & 12.87 & 9.76 \\
    \midrule
    \multirow{9}{*}{LYON}        & NIL & 14.22 & 23.08 & 7.56 & 13.94 & 1.73 & 20.26 & 22.18 & 68.41\\
    \cmidrule{2-10}
                                 & c(90\%) & 13.26 & 21.19 & 6.85 & 12.85 & 1.91 & 19.32 & 21.72 & 60.11\\
                                 & c(80\%) & 13.25 & 20.18 & 6.60 & 12.95 & 1.94 & 18.22 & 20.63 & 56.73\\
                                 & c(70\%) &12.12 & 16.48 & 5.38 & 12.13 &  1.87 & 14.81 & 16.95 & 30.58\\
                                 & c(60\%) & 11.67 & 15.39 & 4.97 & 11.32 & 1.73 & 14.11 & 15.76 & 26.73\\
                                 & c(50\%) & \bf{10.10} & \bf{11.40} & \bf{3.72} & \bf{9.96} & \bf{1.11} & \bf{10.02} & \bf{10.86} & \bf{14.42}\\
    \cmidrule{2-10}
                                 & s(4.0) & 13.96 & 22.74 & 7.50 & 13.80 & 1.76 & 19.94 & 21.85 & 64.71\\
                                 & s(3.0) &13.52 & 21.90 & 7.17 & 13.27 & 1.82 & 19.34 & 21.54 & 60.42\\
                                 & s(2.0) &11.19 & 16.33 & 5.42 & 11.45 & 1.44 & 14.48 & 16.05 & 23.76\\                             
    \midrule
    \multirow{9}{*}{France}      & NIL & 13.61 & 23.90 & 7.15 & 14.04 & 1.83 & 20.52 & 20.87 & 8.11  \\
    \cmidrule{2-10}
                                 & c(90\%) &12.41 & 21.73 & 6.32 & 13.00 & 1.61 & 18.63 & 19.37 & 7.36  \\
                                 & c(80\%) & 11.91 & 20.71 & 5.95 & 12.88 & 1.58 & 17.83 & 18.24 & 7.14\\
                                 & c(70\%) & 11.33 & 19.07 & 5.41 & 12.74 & 1.49 & 16.64 & 17.02 & 6.84\\
                                 & c(60\%) & 9.80 & 16.40 & 4.46 & 10.88 & 1.51 & 14.16 & 14.63 & 5.90\\
                                 & c(50\%) & \bf{8.41} & \bf{13.13} & \bf{3.54} & \bf{9.24} & 1.29 & \bf{11.51} & \bf{11.73} & \bf{4.93}\\
    \cmidrule{2-10}
                                 & s(4.0) & 13.04 & 23.04 & 6.84 & 13.41 & 1.66 & 19.75 & 20.34 & 7.60\\
                                 & s(3.0) & 12.41 & 22.17 & 6.45 & 12.58 & 1.41 & 18.74 & 20.03 & 7.20\\
                                 & s(2.0) & 10.85 & 18.86 & 5.38 & 10.60 & \bf{1.12} & 15.84 & 16.80 & 6.21\\                                 
    \bottomrule
  	\end{tabular}
  	}
  	
\end{table}
\begin{table}[h]
\caption{Percentage of Filtered Data. \label{tab:data_elim}}
\resizebox{0.99\linewidth}{!}
	{
	\begin{tabular}{| l | r | r r r r r |r r r |}
    \toprule
    Filter & NIL & c(90\%) & c(80\%) & c(70\%) & c(60\%)
    & c(50\%) & s(4.0) & s(3.0) & s(2.0)\\
    \midrule
    France\_EHV & 0\% & 12\% & 21\% & 33\% & 44\% & 53\% & 2\% & 7\% & 28\%\\
    \midrule
    LYON & 0\% & 19\% & 37\% & 55\% & 61\% & 70\% & 3\% & 10\% & 48\% \\
    \midrule
    France & 0\% & 16\% & 24\% & 32\% & 45\% & 58\% & 5\% & 11\% & 40\%\\                                  
    \bottomrule
  	\end{tabular}
  	}
  	
\end{table}

\subsection{Learning Accuracy}
%
%
%
Let $T[k]$ to be the collection of the data set obtained by applying
filters to the testing data sets for region $k \in K$.  Let $x^T(t)$
to be the tensor of ground truths for data set $t \in T[k]$, and
$\widehat{x}(t)$ to be predicted tensor.  The mean prediction error
(in \% metric) for $\widehat{x}$ is given by:
\begin{equation}
    100 * \frac{1}{\lvert K \rvert}\sum_{k \in K} \frac{1}{\lvert T[k] \vert } \sum_{t \in T[k]} \frac{\lVert \widehat{x}(t) - x^T(t) \rVert_1}{\lVert x^T(t) \rVert_1}. \nonumber
\end{equation}

Table~\ref{tab:pred_error} 
  and~\ref{tab:pred_error_two_level} (in the appendix for the
  two-level ADMM) present the prediction errors for various filters
for the consensus parameters $\bm{S}^C$, $\bm{V}^C$, and their dual
multipliers $\lambda_{S}$, $\lambda_{V}$.  Since these quantities are
complex numbers, for simplicity, the table presents each individual
component.  Table~\ref{tab:data_elim} also shows the percentage of
instances being filtered.

The results indicate that ADMM solutions are difficult to learn and
generalize.  
For the original ADMM, the errors without filters can be as large as $24\%$ for
primal variables and close to $68\%$ for dual multipliers.
Similar trend can also be observed for the two-level ADMM.
The filters
significantly reduce predictor errors 
for both ADMM variants, producing data sets that are
easier to learn. In particular, 
Table~\ref{tab:pred_error} shows that
the accuracy for primal solutions
(LYON-$\bm{q}^C$) and dual solutions (LYON-$\bm{\lambda}_\theta$)
improve by almost $12\%$ and $54\%$ when using specific filters.

The convergence filters almost always provide higher accuracy than the
standard deviation filters. This may be explained by the fact that the
convergence variations are not necessarily Gaussian and hence the
standard deviation filters are potentially biased against instances
with high infeasibility residues that occur frequently.
\begin{table}[h]
\centering
\caption{Objective Gap Against P-DATA in \% (Avg / \# cases)  \label{tab:quality_measure}}
\resizebox{0.98\linewidth}{!}
	{
	\begin{tabular}{| c | l | r || r  | r | r | r | r | r || r |}
    \toprule
    \multirow{2}{*}{Network}        &  \multirow{2}{*}{Filter} 
    & \multicolumn{8}{c|}{ADMM Iterations}  \\     
    \cmidrule{3-10}
    &   & 5 & 50 & 100 & 150 & 200 & 250 & 300 & 500 \\     
    \midrule
    \multirow{9}{*}{France\_EHV} & NIL &  \bf{-0.17} & \bf{-0.09} & -0.05 & -0.02 & \bf{-0.02} & -0.02 & -0.02 & \bf{0.00} \\ 
                                 & c(90\%)& -0.70 & -0.45 & -0.27 & -0.19 & -0.14 & -0.09 & -0.06 & -0.01 \\  
                                 & c(70\%) & 0.23 & 0.18 & 0.14 & 0.09 & 0.06 & 0.04 & 0.02 & 0.01 \\  
                                 & c(50\%) & 1.10 & 0.73 & 0.42 & 0.25 & 0.17 & 0.10 & 0.06 & 0.02 \\ 
                                 & s(4.0)  & 0.53 & 0.37 & 0.21 & 0.15 & 0.09 & 0.04 & 0.01 & 0.01 \\
                                 & s(3.0)  & 0.59 & 0.34 & 0.16 & 0.08 & 0.03 & \bf{-0.00} & -0.01 & \bf{0.00} \\ 
                                 & s(2.0) & -0.21 & -0.11 & \bf{-0.02} & \bf{0.01} & \bf{0.02} & \bf{0.00} & \bf{-0.00} & 0.01 \\  
                                 \cmidrule{2-10}
                                 & N-ADMM   & -54.32 & -54.08 & -53.99 & -48.37 & -39.09 & -29.64 & -21.23 & -3.44 \\                                   
    \midrule
    \multirow{9}{*}{LYON}        & NIL   & -1.31 & -1.01 & -0.80 & -0.63 & -0.50 & -0.40 & -0.32 & -0.12 \\  
                                 & c(90\%) & \bf{-0.80} & \bf{-0.64} & \bf{-0.50} & \bf{-0.40} & \bf{-0.33} & \bf{-0.26} & \bf{-0.21} & \bf{-0.07} \\
                                 & c(70\%) & -1.47 & -1.18 & -0.91 & -0.73 & -0.59 & -0.48 & -0.40 & -0.20 \\ 
                                 & c(50\%) & -1.29 & -1.04 & -0.83 & -0.66 & -0.54 & -0.45 & -0.37 & -0.18 \\  
                                 & s(4.0) & -1.18 & -0.90 & -0.69 & -0.53 & -0.41 & -0.33 & -0.26 & -0.10 \\
                                 & s(3.0) & -1.20 & -0.94 & -0.73 & -0.58 & -0.46 & -0.37 & -0.30 & -0.12 \\    
                                 & s(2.0) & -0.95 & -0.81 & -0.64 & -0.50 & -0.41 & -0.33 & -0.27 & -0.14 \\ 
                                 \cmidrule{2-10}
                                 & N-ADMM     & -50.42 & -50.04 & -49.27 & -43.73 & -33.84 & -25.90 & -20.46 & -10.44 \\                 
    \midrule
    \multirow{9}{*}{France}      & NIL   & -2.07 & -1.64 & -1.30 & -1.03 & -0.82 & -0.67 & -0.54 & -0.24 \\ 
                                 & c(90\%) &  -2.20 & -1.69 & -1.34 & -1.06 & -0.85 & -0.69 & -0.56 & -0.25 \\  
                                 & c(70\%) &  -1.78 & -1.40 & -1.10 & -0.87 & -0.70 & -0.57 & -0.46 & -0.20 \\  
                                 & c(50\%) &  \bf{-1.27} & \bf{-1.01} & \bf{-0.80} & \bf{-0.63} & \bf{-0.50} & \bf{-0.40} & \bf{-0.31} & \bf{-0.12} \\  
                                 & s(4.0) & -2.12 & -1.65 & -1.31 & -1.04 & -0.83 & -0.68 & -0.55 & -0.23 \\ 
                                 & s(3.0) & -1.71 & -1.36 & -1.07 & -0.84 & -0.67 & -0.53 & -0.42 & -0.17 \\ 
                                 & s(2.0) & -1.64 & -1.30 & -1.02 & -0.81 & -0.63 & -0.49 & -0.39 & -0.15 \\  
                                 \cmidrule{2-10}
                                 & N-ADMM   & -39.99 & -38.43 & -38.38 & -38.38 & -38.38 & -38.17 & -36.73 & -24.21 \\ 
    \bottomrule
  	\end{tabular}
  	}  	
\end{table}
\begin{table}[h]
\centering
\caption{Two-level ADMM: Objective Gap Against P-DATA in \% (Avg/\#)  \label{tab:quality_measure_two_level}}
\resizebox{0.98\linewidth}{!}
	{
	\begin{tabular}{| c | l | r || r  | r | r | r | r | r || r |}
    \toprule
    \multirow{2}{*}{Network}        &  \multirow{2}{*}{Filter} 
    & \multicolumn{8}{c|}{ADMM Iterations}  \\     
    \cmidrule{3-10}
    &   & 5 & 50 & 100 & 150 & 200 & 250 & 300 & 500 \\     
    \midrule
    \multirow{9}{*}{France\_EHV} & NIL &  0.24 & 0.19 & 0.24 & 0.08 & 0.06 & 0.03 & 0.02 & -0.02 \\  
                                 & c(90\%)& \bf{0.23} & \bf{0.16} & \bf{0.21} & \bf{0.06} & \bf{0.05} & \bf{0.02} & 0.02 & \bf{-0.01} \\   
                                 & c(70\%) & 0.38 & 0.32 & 0.38 & 0.12 & 0.10 & 0.06 & 0.05 & 0.02 \\   
                                 & c(50\%) & 0.38 & 0.28 & 0.32 & 0.07 & 0.06 & \bf{0.02} & \bf{0.01} & -0.02 \\  
                                 & s(4.0)  & 0.34 & 0.37 & 0.45 & 0.19 & 0.17 & 0.11 & 0.10 & 0.06 \\ 
                                 & s(3.0)  & 0.40 & 0.35 & 0.43 & 0.18 & 0.16 & 0.11 & 0.10 & 0.06 \\ 
                                 & s(2.0) & 0.37 & 0.28 & 0.36 & 0.12 & 0.10 & 0.07 & 0.06 & 0.04 \\   
                                 \cmidrule{2-10}
                                 & N-ADMM   & -23.01 & -26.43 & -26.52 & -5.59 & -5.65 & -1.08 & -1.10 & -0.07 \\                                    
    \midrule
    \multirow{9}{*}{LYON}        & NIL   & \bf{-0.71} & -0.27 & \bf{0.01} & 0.30 & 0.22 & 0.43 & 0.27 & 1.95 \\   
                                 & c(90\%) & -0.84 & -0.63 & -0.53 & -0.17 & -0.20 & -0.13 & -0.18 & 0.30 \\
                                 & c(70\%) & -1.00 & -0.75 & -0.71 & -0.23 & -0.23 & -0.15 & -0.20 & 0.37 \\ 
                                 & c(50\%) & -0.74 & \bf{0.13} & 0.89 & 0.97 & 0.85 & 1.13 & 0.93 & 4.06 \\   
                                 & s(4.0) & -0.78 & -0.46 & -0.27 & \bf{0.02} & \bf{-0.02} & \bf{0.03} & \bf{-0.05} & 0.51 \\ 
                                 & s(3.0) & -0.87 & -0.64 & -0.55 & -0.26 & -0.29 & -0.24 & -0.28 & \bf{0.18} \\     
                                 & s(2.0) & -0.79 & -0.64 & -0.66 & -0.47 & -0.47 & -0.47 & -0.46 & -0.42 \\  
                                 \cmidrule{2-10}
                                 & N-ADMM & -24.44 & -30.03 & -30.26 & -8.15 & -8.15 & -2.06 & -2.02 & 0.27 \\                  
    \midrule
    \multirow{9}{*}{France}      & NIL   & 0.03 & 0.06 & 0.12 & 0.03 & \bf{0.00} & -0.09 & -0.13 & \bf{-0.10} \\  
                                 & c(90\%) &  0.12 & 0.09 & 0.12 & -0.08 & -0.09 & -0.17 & -0.17 & -0.22 \\  
                                 & c(70\%) &  0.05 & -0.09 & \bf{-0.06} & -0.10 & -0.12 & -0.14 & -0.14 & -0.17 \\   
                                 & c(50\%) &  -0.17 & -0.34 & -0.31 & -0.18 & -0.19 & -0.15 & -0.15 & -0.15 \\   
                                 & s(4.0) & \bf{0.02} & \bf{0.04} & 0.10 & 0.01 & -0.03 & -0.10 & -0.14 & -0.15 \\  
                                 & s(3.0) & 0.05 & 0.07 & 0.13 & \bf{0.00} & -0.02 & -0.11 & -0.15 & -0.19 \\  
                                 & s(2.0) & 0.41 & 0.50 & 0.64 & 0.26 & 0.22 & \bf{0.08} & \bf{0.04} & 0.13 \\  
                                 \cmidrule{2-10}
                                 & N-ADMM   & -29.00 & -34.99 & -35.16 & -10.99 & -10.97 & -2.24 & -2.33 & -0.10 \\  
    \bottomrule
  	\end{tabular}
  	}  	
\end{table}
\subsection{Performance of ML-ADMM Against the Ground Truth}

This section presents the performance results of ML-ADMM over
\textit{\textbf{all testing instances}}, including those instances
with high infeasibility residues.  Table~\ref{tab:quality_measure}
and~\ref{tab:quality_measure_two_level} 
present the average objective gap of ML-ADMM (over all test cases and
regions $k \in K$) against the ground truth P-DATA when ML-ADMM is run
for a number of iterations ranging from 5 to 500 ADMM.  The average
objective gaps of N-ADMM are also included for comparison purposes.
Let ${\cO}^{*}$ to be the objective value from P-DATA and let
$\hat{\cO}$ be the objective value returned by a run of ML-ADMM. The
  objective gap is defined as
\begin{equation}
     100 \times \frac{\hat{\cO} - {\cO}^{*}}{{\cO}^{*}} \nonumber
\end{equation}
In addition, Tables~\ref{tab:residue_iter} and~\ref{tab:residue_iter2}
(and Tables~\ref{tab:residue_iter_two_level} 
and~\ref{tab:residue_iter2_two_level} for the two-level ADMM)
also report average results for the primal and dual residues
$\bm{r}_{p}$ and $\bm{r}_{d}$.

The results show that ML-ADMM provides orders of magnitude
improvements in objective gap, primal residue, and dual residue over
ADMM for small numbers of iterations. Within 500 iterations, the
ML-ADMM variants for the original ADMM recover almost
the same solution quality as P-DATA ($< 0.3\%$ objective difference).
For the two-level ADMM, ML-ADMM variants recovers almost
  the same solution quality in less than 200 iterations, which is
  natural since the two-level ADMM generally converges faster than the
  original variant.  Interestingly, within 5 iterations, the ML-ADMM
variants differ only by at most $3\%$ (1\% for the
  two-level ADMM) from the ground truth. On the contrary, N-ADMM
exhibits objective gaps of $-3.44\%$, $-10.44\%$, and $-24.21\%$
($-23.01\%, -24.44\%,$ and $-29\%$ for the two-level
  ADMM), demonstrating the value of learning for fast convergence.

\begin{table}[t]
\centering
\caption{Primal Infeas. Residue in p.u. (Avg / \# cases)  \label{tab:residue_iter}}
\resizebox{0.70\linewidth}{!}
	{
	\begin{tabular}{| c | r| r | r  | r | r | r | r |}
    \toprule
    \multirow{2}{*}{Network}        &  \multirow{2}{*}{Start/Init.} 
    & \multicolumn{6}{c|}{ADMM Iterations}  \\     
    \cmidrule{3-8}
    &   & 50 & 100 & 150 & 200 & 250 & 300 \\     
    \midrule
    \multirow{9}{*}{France\_EHV} & NIL&   0.10 & 0.06 & 0.04 & 0.03 & 0.02 & 0.02\\
                                 & c(90\%)&  0.09 & 0.06 & 0.04 & 0.03 & 0.02 & 0.02\\  
                                 & c(70\%) & 0.10 & 0.06 & 0.04 & 0.03 & 0.03 & 0.02\\ 
                                 & c(50\%) & 0.11 & 0.07 & 0.05 & 0.03 & 0.03 & 0.02\\
                                 & s(4.0)  & 0.10 & 0.06 & 0.04 & 0.03 & 0.02 & 0.02\\
                                 & s(3.0)  & 0.09 & 0.06 & 0.04 & 0.03 & 0.02 & 0.02\\ 
                                 & s(2.0) & 0.11 & 0.07 & 0.05 & 0.04 & 0.03 & 0.02\\ 
                                 \cmidrule{2-8}
                                 & N-ADMM       & 2.24 & 2.22 & 1.82 & 1.51 & 1.34 & 1.38\\ 
                                 \cmidrule{2-8}
                                 & P-ADMM       & 0.02 & 0.02 & 0.02 & 0.02 & 0.02 & 0.02\\ 
                                
    \midrule
    \multirow{9}{*}{LYON}        & NIL  & 0.18 & 0.17 & 0.17 & 0.16 & 0.16 & 0.15\\ 
                                 & c(90\%) & 0.18 & 0.17 & 0.17 & 0.16 & 0.16 & 0.15\\ 
                                 & c(70\%) & 0.18 & 0.18 & 0.17 & 0.17 & 0.16 & 0.16\\ 
                                 & c(50\%) & 0.19 & 0.19 & 0.18 & 0.18 & 0.18 & 0.17\\ 
                                 & s(4.0) & 0.18 & 0.17 & 0.17 & 0.16 & 0.16 & 0.15\\ 
                                 & s(3.0) & 0.18 & 0.17 & 0.16 & 0.16 & 0.16 & 0.15\\   
                                 & s(2.0) & 0.19 & 0.19 & 0.18 & 0.18 & 0.17 & 0.17\\
                                 \cmidrule{2-8}
                                 & N-ADMM         & 2.10 & 1.91 & 1.62 & 1.4 & 1.18 & 1.00\\ 
                                 \cmidrule{2-8}
                                 & P-ADMM         & 0.08 & 0.08 & 0.08 & 0.08 & 0.08 & 0.07\\ 
    \midrule
    \multirow{9}{*}{France}      & NIL &  0.08 & 0.07 & 0.06 & 0.07 & 0.07 & 0.07\\ 
                                 & c(90\%) &  0.08 & 0.07 & 0.06 & 0.06 & 0.07 & 0.06\\ 
                                 & c(70\%) &  0.08 & 0.07 & 0.06 & 0.06 & 0.06 & 0.07\\ 
                                 & c(50\%) &  0.07 & 0.06 & 0.06 & 0.06 & 0.06 & 0.06\\ 
                                 & s(4.0) & 0.08 & 0.07 & 0.06 & 0.06 & 0.07 & 0.07\\
                                 & s(3.0) & 0.08 & 0.07 & 0.06 & 0.07 & 0.06 & 0.07\\
                                 & s(2.0) & 0.08 & 0.07 & 0.06 & 0.06 & 0.06 & 0.07\\ 
                                 \cmidrule{2-8}
                                 & N-ADMM   &  0.80 & 0.75 & 0.74 & 0.74 & 0.73 & 0.69\\
                                 \cmidrule{2-8}
                                 & P-ADMM         & 0.08 & 0.09 & 0.09 & 0.08 & 0.09 & 0.09\\
    \bottomrule
  	\end{tabular}
  	}  	
\end{table}
\begin{table}[t]
\centering
\caption{Two-level ADMM: Primal Infeas. Residue in p.u. (Avg / \# cases)   
\label{tab:residue_iter_two_level}}
\resizebox{0.70\linewidth}{!}
	{
	\begin{tabular}{| c | r| r | r  | r | r | r | r |}
    \toprule
    \multirow{2}{*}{Network}        &  \multirow{2}{*}{Start/Init.} 
    & \multicolumn{6}{c|}{ADMM Iterations}  \\     
    \cmidrule{3-8}
    &   & 50 & 100 & 150 & 200 & 250 & 300 \\     
    \midrule
    \multirow{9}{*}{France\_EHV} & NIL&   0.09 & 0.12 & 0.10 & 0.09 & 0.08 & 0.08\\ 
                                 & c(90\%)&  0.09 & 0.12 & 0.11 & 0.11 & 0.10 & 0.10\\   
                                 & c(70\%) & 0.09 & 0.14 & 0.13 & 0.13 & 0.13 & 0.12\\  
                                 & c(50\%) & 0.08 & 0.10 & 0.06 & 0.06 & 0.06 & 0.05\\ 
                                 & s(4.0)  & 0.09 & 0.12 & 0.12 & 0.12 & 0.12 & 0.12\\
                                 & s(3.0)  & 0.09 & 0.13 & 0.13 & 0.13 & 0.14 & 0.13\\  
                                 & s(2.0) & 0.09 & 0.11 & 0.08 & 0.07 & 0.07 & 0.07\\  
                                 \cmidrule{2-8}
                                 & N-ADMM   & 0.89 & 0.89 & 0.17 & 0.17 & 0.07 & 0.06\\ 
                                 \cmidrule{2-8}
                                 & P-ADMM   & 0.06 & 0.09 & 0.08 & 0.10 & 0.07 & 0.07\\ 
                                
    \midrule
    \multirow{9}{*}{LYON}        & NIL  & 0.31 & 0.43 & 0.36 & 0.34 & 0.23 & 0.21\\  
                                 & c(90\%) & 0.18 & 0.27 & 0.24 & 0.22 & 0.17 & 0.15\\  
                                 & c(70\%) & 0.15 & 0.26 & 0.24 & 0.22 & 0.18 & 0.17\\ 
                                 & c(50\%) & 0.36 & 0.56 & 0.52 & 0.50 & 0.36 & 0.31\\  
                                 & s(4.0) & 0.24 & 0.32 & 0.29 & 0.28 & 0.20 & 0.18\\  
                                 & s(3.0) & 0.17 & 0.23 & 0.21 & 0.2 & 0.15 & 0.14\\    
                                 & s(2.0) & 0.12 & 0.14 & 0.13 & 0.12 & 0.11 & 0.10\\ 
                                 \cmidrule{2-8}
                                 & N-ADMM    & 0.88 & 0.86 & 0.21 & 0.19 & 0.16 & 0.14\\ 
                                 \cmidrule{2-8}
                                 & P-ADMM    &0.04 & 0.07 & 0.05 & 0.05 & 0.04 & 0.04\\ 
    \midrule
    \multirow{9}{*}{France}      & NIL &  0.31 & 0.79 & 0.55 & 0.51 & 0.32 & 0.22\\  
                                 & c(90\%) &  0.09 & 0.13 & 0.07 & 0.08 & 0.05 & 0.07\\  
                                 & c(70\%) &  0.09 & 0.13 & 0.09 & 0.07 & 0.06 & 0.06\\ 
                                 & c(50\%) &  0.08 & 0.12 & 0.10 & 0.10 & 0.09 & 0.08\\  
                                 & s(4.0) & 0.24 & 0.75 & 0.50 & 0.42 & 0.27 & 0.18\\ 
                                 & s(3.0) & 0.16 & 0.56 & 0.40 & 0.38 & 0.24 & 0.17\\ 
                                 & s(2.0) & 0.66 & 0.96 & 0.58 & 0.60 & 0.41 & 0.30\\  
                                 \cmidrule{2-8}
                                 & N-ADMM   &  0.62 & 0.62 & 0.15 & 0.14 & 0.09 & 0.08\\ 
                                 \cmidrule{2-8}
                                 & P-ADMM   & 1.78 & 1.83 & 0.34 & 0.36 & 0.26 & 0.26\\ 
    \bottomrule
  	\end{tabular}
  	}  	
\end{table}
\begin{table}[h]
\centering
\caption{Dual Infeas. Residue in p.u. (Avg / \# cases)  \label{tab:residue_iter2}}
\resizebox{0.70\linewidth}{!}
	{
	\begin{tabular}{| c | r| r | r  | r | r | r | r |}
    \toprule
    \multirow{2}{*}{Network}        &  \multirow{2}{*}{Start/Init.} 
    & \multicolumn{6}{c|}{ADMM Iterations}  \\     
    \cmidrule{3-8}
    &   & 50 & 100 & 150 & 200 & 250 & 300 \\     
    \midrule
    \multirow{9}{*}{France\_EHV} & NIL &    0.12 & 0.07 & 0.04 & 0.03 & 0.02 & 0.01\\ 
                                 & c(90\%)&   0.12 & 0.07 & 0.05 & 0.03 & 0.02 & 0.01\\
                                 & c(70\%) &  0.13 & 0.08 & 0.05 & 0.03 & 0.02 & 0.01\\
                                 & c(50\%) &  0.13 & 0.07 & 0.05 & 0.03 & 0.02 & 0.01\\
                                 & s(4.0)  &  0.11 & 0.06 & 0.04 & 0.03 & 0.02 & 0.01\\ 
                                 & s(3.0)  &  0.13 & 0.07 & 0.05 & 0.03 & 0.02 & 0.01\\ 
                                 & s(2.0) &   0.14 & 0.08 & 0.05 & 0.03 & 0.02 & 0.01\\ 
                                 \cmidrule{2-8}
                                 & N-ADMM       & 0.17 & 0.27 & 1.47 & 1.29 & 1.17 & 1.01\\ 
                                 \cmidrule{2-8}
                                 & P-ADMM       & 0.00 & 0.00 & 0.00 & 0.00 & 0.00 & 0.00\\ 
                                
    \midrule
    \multirow{9}{*}{LYON}        & NIL & 0.10 & 0.07 & 0.11 & 0.14 & 0.12 & 0.18\\ 
                                 & c(90\%) & 0.09 & 0.06 & 0.09 & 0.11 & 0.15 & 0.16\\ 
                                 & c(70\%) & 0.16 & 0.18 & 0.17 & 0.24 & 0.24 & 0.21\\ 
                                 & c(50\%) & 0.23 & 0.17 & 0.24 & 0.23 & 0.27 & 0.21\\ 
                                 & s(4.0) & 0.10 & 0.07 & 0.11 & 0.20 & 0.14 & 0.14\\
                                 & s(3.0) & 0.09 & 0.07 & 0.11 & 0.15 & 0.16 & 0.15\\   
                                 & s(2.0) & 0.19 & 0.18 & 0.19 & 0.26 & 0.25 & 0.22\\
                                 \cmidrule{2-8}
                                 & N-ADMM   & 0.42 & 0.62 & 1.24 & 1.03 & 0.76 & 0.66\\
                                 \cmidrule{2-8}
                                 & P-ADMM   & 0.25 & 0.27 & 0.25 & 0.22 & 0.25 & 0.24\\
    \midrule
    \multirow{9}{*}{France}      & NIL &  0.05 & 0.03 & 0.04 & 0.11 & 0.10 & 0.17\\
                                 & c(90\%) &  0.05 & 0.03 & 0.03 & 0.06 & 0.12 & 0.13\\ 
                                 & c(70\%) &  0.04 & 0.03 & 0.03 & 0.05 & 0.09 & 0.14\\
                                 & c(50\%) &  0.04 & 0.02 & 0.03 & 0.04 & 0.12 & 0.16\\ 
                                 & s(4.0) &   0.05 & 0.03 & 0.04 & 0.08 & 0.14 & 0.19\\ 
                                 & s(3.0) &   0.04 & 0.02 & 0.02 & 0.10 & 0.12 & 0.16\\ 
                                 & s(2.0) &   0.05 & 0.03 & 0.02 & 0.04 & 0.07 & 0.15\\ 
                                 \cmidrule{2-8}
                                 & N-ADMM   &   0.53 & 0.18 & 0.12 & 0.09 & 0.2 & 0.39\\ 
                                 \cmidrule{2-8}
                                 & P-ADMM    &  0.45 & 0.54 & 0.47 & 0.47 & 0.54 & 0.50\\ 
    \bottomrule
  	\end{tabular}
  	}  	
\end{table}

\begin{table}[h]
\centering
\caption{Two-level ADMM: Dual Infeas. Residue 
in p.u. (Avg / \# cases)   
\label{tab:residue_iter2_two_level}}
\resizebox{0.80\linewidth}{!}
	{
	\begin{tabular}{| c | r| r | r  | r | r | r | r |}
    \toprule
    \multirow{2}{*}{Network}        &  \multirow{2}{*}{Start/Init.} 
    & \multicolumn{6}{c|}{ADMM Iterations}  \\     
    \cmidrule{3-8}
    &   & 50 & 100 & 150 & 200 & 250 & 300 \\     
    \midrule
    \multirow{9}{*}{France\_EHV} & NIL     & 30.29 & 23.72 & 6.59 & 5.69 & 5.14 & 6.31\\ 
                                 & c(90\%) & 38.86 & 31.68 & 7.14 & 7.78 & 6.89 & 9.24\\  
                                 & c(70\%) & 52.68 & 43.48 & 5.47 & 4.53 & 3.20 & 2.73\\  
                                 & c(50\%) & 21.34 & 17.14 & 7.53 & 6.99 & 4.36 & 4.80\\
                                 & s(4.0)  & 58.81 & 49.66 & 5.63 & 4.83 & 3.73 & 5.94\\ 
                                 & s(3.0)  & 62.70 & 52.19 & 5.64 & 4.66 & 3.37 & 6.08\\ 
                                 & s(2.0)  & 27.63 & 24.45 & 6.61 & 5.14 & 3.79 & 3.72\\ 
                                 \cmidrule{2-8}
                                 & N-ADMM  & 32.35 & 74.19 & 117.84 & 313.35 & 152.99 & 123.49\\ 
                                 \cmidrule{2-8}
                                 & P-ADMM  & 12.82 & 19.57 & 15.26 & 20.24 & 17.28 & 40.15\\  
                                
    \midrule
    \multirow{9}{*}{LYON}        & NIL  & 74.85 & 287.27 & 66.91 & 762.44 & 548.26 & 6003.08\\ 
                                 & c(90\%) & 51.51 & 130.48 & 34.06 & 222.29 & 186.66 & 1933.53\\ 
                                 & c(70\%) & 55.12 & 128.15 & 31.95 & 141.64 & 174.53 & 854.69\\ 
                                 & c(50\%) & 125.78 & 320.95 & 70.35 & 847.05 & 671.58 & 4763.11\\  
                                 & s(4.0) & 59.41 & 221.72 & 43.90 & 412.77 & 357.19 & 2879.60\\  
                                 & s(3.0) & 50.10 & 134.88 & 29.80 & 165.56 & 125.89 & 874.56\\   
                                 & s(2.0) & 34.25 & 54.86 & 18.21 & 82.30 & 54.13 & 285.85\\ 
                                 \cmidrule{2-8}
                                 & N-ADMM    & 61.74 & 198.98 & 98.64 & 330.94 & 351.90 & 892.89\\ 
                                 \cmidrule{2-8}
                                 & P-ADMM    &14.58 & 25.63 & 13.56 & 29.03 & 13.74 & 94.47\\  
    \midrule
    \multirow{9}{*}{France}      & NIL     &  73.38 & 383.95 & 101.69 & 552.82 & 876.88 & 3330.52\\  
                                 & c(90\%) &  21.18 & 18.42 & 15.89 & 21.70 & 12.51 & 22.77\\  
                                 & c(70\%) &  22.23 & 16.52 & 9.26 & 13.78 & 3.98 & 7.49\\  
                                 & c(50\%) &  26.46 & 18.81 & 5.83 & 6.01 & 4.42 & 5.09\\  
                                 & s(4.0) & 71.90 & 359.84 & 102.93 & 971.85 & 801.91 & 940.26\\
                                 & s(3.0) & 50.68 & 537.00 & 81.42 & 321.68 & 493.82 & 385.94\\
                                 & s(2.0) & 166.22 & 174.47 & 144.77 & 905.70 & 877.01 & 5112.71\\  
                                 \cmidrule{2-8}
                                 & N-ADMM   &  35.95 & 105.54 & 122.47 & 345.92 & 249.12 & 414.81\\ 
                                 \cmidrule{2-8}
                                 & P-ADMM   & 92.81 & 265.51 & 145.11 & 1739.56 & 523.89 & 6048.51\\ 
    \bottomrule
  	\end{tabular}
  	}  	
\end{table}

The primal and dual residue of the ML-ADMM variants are of high
quality, often improving those of P-ADMM, the ADMM procedure
initialized with the ground-truth consensus and dual values. This
contrast with the results of N-ADMM which are often an order magnitude
larger than those of the ML-ADMM variants.  The
  two-level ADMM exhibits a larger magnitude of dual infeasibility
  residue due to its two-level structure and its generally more
  aggressive convergence nature (resulting in more significant swings
  between iterations).  With machine learning, it is worth to note
  that ML-ADMM may stabilize the search even faster than P-ADMM,
  resulting in a significantally smaller dual infeasibility residue.

These results indicate that hot-starting the ADMM procedure with
machine-learning predictions is effective in producing solutions of
the same quality as the traditional ADMM procedure with a fraction of
the number of iterations. In particular, 500 iterations of ML-ADMM is
essentially similar to 3,000 iterations of the standard ADMM. ML-ADMM
with even fewer iterations provides high-quality approximations to the
traditional ADMM. 
Applying filters improves the quality of ML-ADMM.

\begin{table}[t]
\centering
\caption{Average Optimality Gap (\%) against Centralized Routine (\%) \label{tab:centralized_measure}}
\resizebox{0.90\linewidth}{!}
	{
	\begin{tabular}{| c | r| r | r  | r | r | r | r |}
    \toprule
    \multirow{2}{*}{Network}        &  \multirow{2}{*}{Start/Init.} 
    & \multicolumn{6}{c|}{ADMM Iterations}  \\     
    \cmidrule{3-8}
    &   & 50 & 100 & 150 & 200 & 250 & 300 \\     
    \midrule
    \multirow{9}{*}{France\_EHV} & ML-ADMM &  -0.27 & -0.23 & -0.20 & -0.20 & -0.20 & -0.19\\
                                 & + c(90\%)& -0.63 & -0.45 & -0.37 & -0.32 & -0.27 & -0.24\\  
                                 & + c(70\%) &\bf{-0.01} & -0.04 & -0.09 & -0.12 & -0.14 & -0.16\\  
                                 & + c(50\%) &0.54 & 0.24 & 0.07 & \bf{-0.02} & \bf{-0.08} & \bf{-0.12}\\
                                 & + s(4.0)  &0.18 & 0.03 & \bf{-0.03} & -0.09 & -0.14 & -0.17\\
                                 & + s(3.0)  &0.16 & \bf{-0.02} & -0.10 & -0.15 & -0.18 & -0.19\\
                                 & + s(2.0) & -0.29 & -0.20 & -0.17 & -0.16 & -0.18 & -0.18\\  
                                 \cmidrule{2-8}
                                 & N-ADMM       & -54.15 & -54.06 & -48.46 & -39.20 & -29.77 & -21.37\\  
                                 \cmidrule{2-8}
                                 & P-ADMM       & -0.18 & -0.18 & -0.18 & -0.18 & -0.18 & -0.18\\
                                
    \midrule
    \multirow{9}{*}{LYON}        & ML-ADMM   & -1.52 & -1.32 & -1.15 & -1.02 & -0.92 & -0.84\\
                                 & + c(90\%) & \bf{-1.16} & \bf{-1.02} & \bf{-0.92} & \bf{-0.85} & \bf{-0.79} & \bf{-0.73}\\
                                 & + c(70\%) & -1.69 & -1.43 & -1.25 & -1.11 & -1.00 & -0.92\\  
                                 & + c(50\%) & -1.56 & -1.34 & -1.18 & -1.06 & -0.97 & -0.89\\  
                                 & + s(4.0) & -1.41 & -1.21 & -1.05 & -0.93 & -0.85 & -0.79\\  
                                 & + s(3.0) & -1.45 & -1.25 & -1.10 & -0.99 & -0.90 & -0.82\\  
                                 & + s(2.0) & -1.33 & -1.16 & -1.02 & -0.93 & -0.85 & -0.80\\
                                 \cmidrule{2-8}
                                 & N-ADMM         & -50.28 & -49.52 & -44.03 & -34.19 & -26.28 & -20.88\\  
                                 \cmidrule{2-8}
                                 & P-ADMM         & -0.52 & -0.51 & -0.50 & -0.49 & -0.49 & -0.48\\         
    \midrule
    \multirow{9}{*}{France}      & ML-ADMM   & -2.90 & -2.57 & -2.30 & -2.10 & -1.95 & -1.83\\  
                                 & + c(90\%) & -2.96 & -2.61 & -2.34 & -2.14 & -1.98 & -1.86\\
                                 & + c(70\%) & -2.66 & -2.37 & -2.15 & -1.98 & -1.85 & -1.75\\ 
                                 & + c(50\%) & \bf{-2.27} & \bf{-2.07} & \bf{-1.91} & \bf{-1.79} & \bf{-1.68} & \bf{-1.60}\\  
                                 & + s(4.0) & -2.91 & -2.58 & -2.31 & -2.11 & -1.96 & -1.83\\  
                                 & + s(3.0) & -2.63 & -2.34 & -2.12 & -1.95 & -1.82 & -1.71\\   
                                 & + s(2.0) & -2.56 & -2.29 & -2.08 & -1.92 & -1.78 & -1.67\\  
                                 \cmidrule{2-8}
                                 & N-ADMM   & -39.31 & -39.27 & -39.26 & -39.26 & -39.06 & -37.63\\  
                                 \cmidrule{2-8}
                                 & P-ADMM         & -1.30 & -1.29 & -1.29 & -1.28 & -1.28 & -1.28\\
    \bottomrule
  	\end{tabular}
  	}  	
\end{table}

\begin{table}[t]
\centering
\caption{Two-level ADMM: Average Optimality Gap (\%) against Centralized Routine (\%) \label{tab:centralized_measure_two_level}}
\resizebox{0.90\linewidth}{!}
	{
	\begin{tabular}{| c | r| r | r  | r | r | r | r |}
    \toprule
    \multirow{2}{*}{Network}        &  \multirow{2}{*}{Start/Init.} 
    & \multicolumn{6}{c|}{ADMM Iterations}  \\     
    \cmidrule{3-8}
    &   & 50 & 100 & 150 & 200 & 250 & 300 \\     
    \midrule
    \multirow{9}{*}{France\_EHV} & ML-ADMM  & 0.21 & 0.26 & 0.11 & 0.09 & 0.05 & \bf{0.04}\\
                                 & + c(90\%)& \bf{0.18} & \bf{0.24} & \bf{0.09} & \bf{0.08} & \bf{0.04} & \bf{0.04}\\   
                                 & + c(70\%) &0.35 & 0.40 & 0.14 & 0.13 & 0.08 & 0.08\\  
                                 & + c(50\%) &0.30 & 0.34 & 0.10 & 0.09 & \bf{0.04} & \bf{0.04}\\ 
                                 & + s(4.0)  &0.39 & 0.47 & 0.21 & 0.19 & 0.13 & 0.12\\ 
                                 & + s(3.0)  &0.38 & 0.45 & 0.20 & 0.19 & 0.13 & 0.12\\ 
                                 & + s(2.0) & 0.30 & 0.38 & 0.14 & 0.13 & 0.09 & 0.09\\   
                                 \cmidrule{2-8}
                                 & N-ADMM   & -26.41 & -26.50 & -5.57 & -5.63 & -1.06 & -1.07\\   
                                 \cmidrule{2-8}
                                 & P-ADMM     & 0.03 & 0.05 & 0.04 & 0.05 & 0.04 & 0.03\\
                                
    \midrule
    \multirow{9}{*}{LYON}        & ML-ADMM   & \bf{0.01} & 0.29 & 0.57 & 0.50 & 0.71 & 0.55\\
                                 & + c(90\%) & -0.36 & -0.25 & 0.11 & 0.07 & 0.14 & 0.10\\
                                 & + c(70\%) & -0.48 & -0.44 & 0.05 & 0.05 & 0.13 & 0.08\\  
                                 & + c(50\%) & 0.41 & 1.17 & 1.25 & 1.13 & 1.41 & 1.21\\   
                                 & + s(4.0) & -0.19 & \bf{0.01} & 0.30 & 0.25 & 0.31 & 0.23\\  
                                 & + s(3.0) & -0.36 & -0.28 & \bf{0.01} & \bf{-0.01} & \bf{0.04} & \bf{0.00}\\  
                                 & + s(2.0) & -0.37 & -0.38 & -0.19 & -0.20 & -0.19 & -0.19\\ 
                                 \cmidrule{2-8}
                                 & N-ADMM     & -29.84 & -30.07 & -7.89 & -7.89 & -1.79 & -1.75\\   
                                 \cmidrule{2-8}
                                 & P-ADMM     & 0.00 & 0.04 & 0.07 & 0.04 & 0.05 & 0.04\\         
    \midrule
    \multirow{9}{*}{France}      & ML-ADMM   & -0.47 & -0.41 & -0.49 & -0.52 & -0.61 & -0.65\\   
                                 & + c(90\%) & -0.44 & -0.41 & -0.60 & -0.62 & -0.69 & -0.69\\ 
                                 & + c(70\%) & -0.62 & -0.59 & -0.62 & -0.64 & -0.66 & -0.66\\  
                                 & + c(50\%) & -0.87 & -0.84 & -0.70 & -0.71 & -0.67 & -0.67\\   
                                 & + s(4.0) & -0.48 & -0.43 & -0.52 & -0.56 & -0.62 & -0.66\\   
                                 & + s(3.0) & -0.46 & -0.40 & -0.52 & -0.55 & -0.64 & -0.67\\   
                                 & + s(2.0) & \bf{-0.03} & \bf{0.11} & \bf{-0.26} & \bf{-0.31} & \bf{-0.45} & \bf{-0.48}\\   
                                 \cmidrule{2-8}
                                 & N-ADMM   & -35.38 & -35.55 & -11.46 & -11.44 & -2.75 & -2.84\\   
                                 \cmidrule{2-8}
                                 & P-ADMM     & -0.67 & -0.66 & -0.38 & -0.38 & -0.27 & -0.25\\ 
    \bottomrule
  	\end{tabular}
  	}  	
\end{table}
\begin{figure}[t]
\centering
 \includegraphics[width=0.70\linewidth]{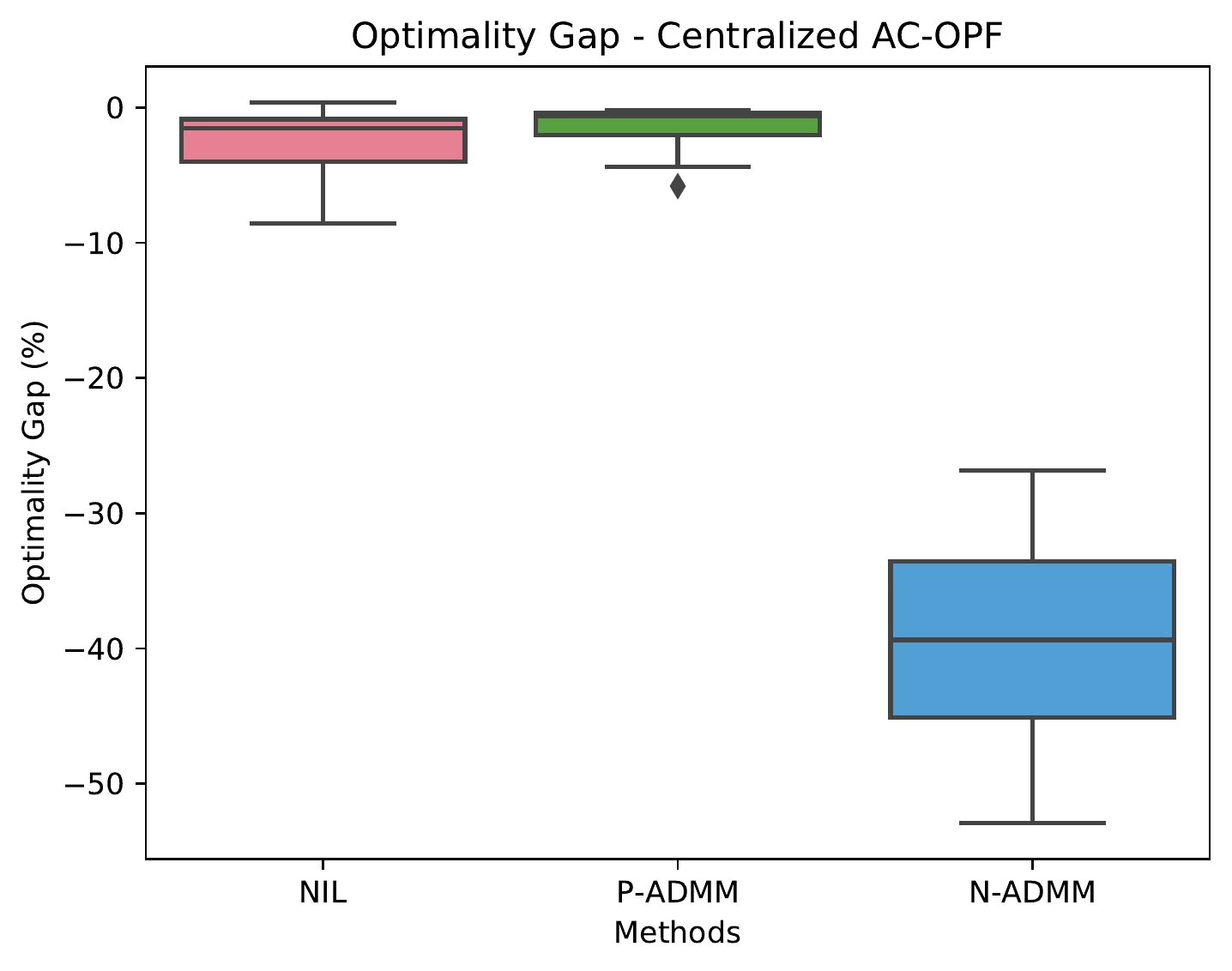}
 \includegraphics[width=0.70\linewidth]{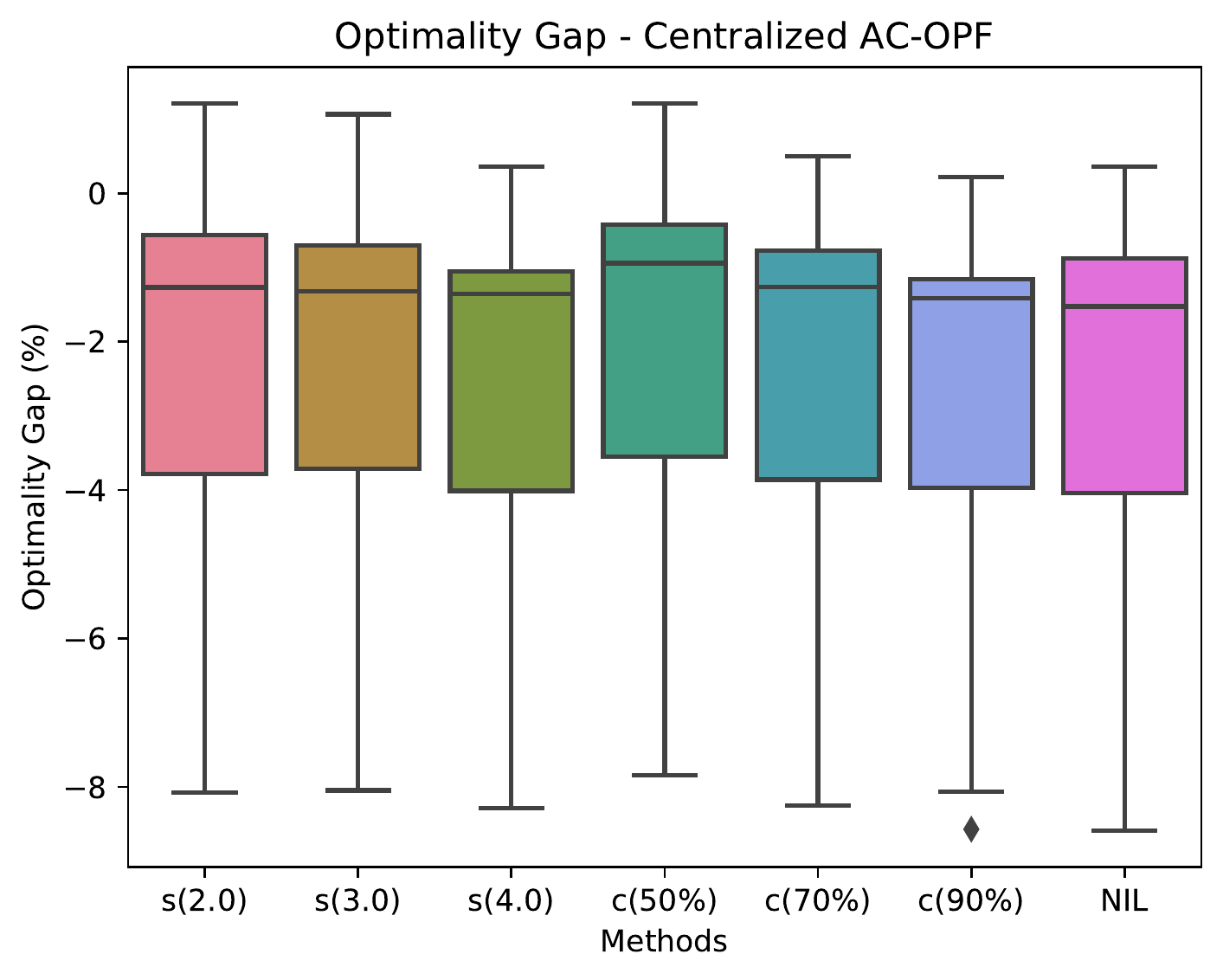}
\caption{Optimality Gap Box Plots, on France, at 100 ADMM iterations.}
\label{fig:box_plots}
\end{figure}

\subsection{Performance of ML-ADMM Against Centralized AC-OPF}

Since not all the test cases from P-DATA converged with the same level
of infeasibility, the comparison against the ground-truths may not
always be ideal: indeed, ML-ADMM may obtain solutions with potentially
better objectives, but these would be reported as errors in
Tables ~\ref{tab:quality_measure} and \ref{tab:quality_measure_two_level}.
This section further evaluates the
learning routines against solutions obtained by a centralized AC-OPF
procedure, which almost always produces better objective values.


Table~\ref{tab:centralized_measure} and~\ref{tab:centralized_measure_two_level}
report the average optimality gap
over all the testing instances.  Again, the ML-ADMM variants provide
orders of magnitude improvements in optimality gaps over N-ADMM.
For the two-level ADMM with faster convergence, ML-ADMM 
can even produce solutions within $1\%$ optimality gap
against centralized AC-OPF routine with only 50 ADMM iterations. 
Moreover, the ML-ADMM
variants recover almost the same optimality gap as P-ADMM and 
could even deliver
a smaller optimality gap for the France\_EHV benchmark 
(LYON for the two-level ADMM).
Again, filters
further improve the benefits of learning.  As indicated in
previous tables,
P-ADMM (seeded with P-DATA) have
almost no improvements.

Figure~\ref{fig:box_plots} further shows the optimality gap statistics
(summarized by two box(-and-whisker) plots) for all the testing
instances for the largest French benchmark for the traditional ADMM.
The box-plots indicate that
ML-ADMM is essentially similar to P-ADMM, which is initialized with
the ground truth, and produces orders of magnitude improvements
compared to the traditional ADMM.
The
box-plots indicate tighter filtering parameters would generally result
in slightly better skew/median optimality gaps, and with largely
similar spread and variance.

\section{Conclusion}
\label{sec:conclusion}

This paper proposed ML-ADMM, a decentralized machine-learning framework
to accelerate the convergence of an ADMM algorithm for solving the
AC-OPF problem. The framework learns the coupling parameters of the
regionally decentralized AC-OPF formulation, which can be used to hot-start the ADMM algorithm when
new instances arrived. The paper has also explored the benefits of 
learning filters --- filters that prevent 
machine learning being trained 
on instances with bad convergence properties. Experimental results on
data sets from the French networks have showed that ML-ADMM produces
solutions of similar quality than the traditional 
and two-level ADMM algorithms
within a fraction (1/6) of iterations (500 versus 3,000). Moreover,
ML-ADMM can produce solutions of simlar quality as the ADMM algorithm
hot-started with the ground truths for the consensus and dual
multipliers. Filtering the datasets to learn from ``good'' runs
also generally provides some additional benefits. These results indicate that machine
learning could be a valuable tool for future smart grids operated with
distributed optimization algorithms
similar to ADMM. 

\section*{Acknowledgments}

This research is partly funded by NSF Awards 2007095 and 2112533.

\bibliographystyle{IEEEtran}
\bibliography{dl_opf,dp}



\appendices
\section*{Appendix}

\begin{model}[h]
	{\small
	\caption{Inner-Level Augmented Lagrangian: $P_{L}$}
	\label{model:two-level_admm_ac_opf}
	\vspace{-6pt}
	\begin{align}
        \mbox{\bf input: } \;\; & \bm{S^d}[k] = ( S^d_i: i\in N_k) \nonumber \\
        & \bm{S^C}[k] = (S^C_{ij} : (i,j) \in R_k \cup R_k^R) \nonumber \\
        & \bm{\lambda_S}[k] = (\lambda_{S^f_{ij}} : (i,j) \in R_k \cup R_k^R) \nonumber \\
        & \bm{z_S}[k] = (z_{S^f_{ij}} : (i,j) \in R_k \cup R_k^R) \nonumber \\
        & \bm{V^C}[k] = (V^C_{i} : (i,j) \in R_k \cup R_k^R) \nonumber \\
        & \bm{\lambda_V}[k] = (\lambda_{V_{i}} : (i,j) \in R_k \cup R_k^R) \nonumber \\
        & \bm{z_V}[k] = (z_{V_{i}} : (i,j) \in R_k \cup R_k^R) \nonumber \\
        & \rho \nonumber \\  	
		\mbox{\bf variables: } \;\;
		& \bm{S^g}[k] =(S^g_i: \forall i \in N_k),  \nonumber \\
		& \bm{V}[k] = (V_i: \forall i \in N_k \cup N^N_k) \nonumber \\
		& {\bm{S^f}[k] = (S^f_{ij} :  \forall(i,j)\in E_k \cup E_k^R \cup R_k \cup R_k^R )} \nonumber \\  	
		\mbox{\bf minimize: } \;\;
		& \sum_{i \in N_k} M_{i}(\Re(S^g_i)) +\nonumber \\
		&  \mkern-25mu \displaystyle\sum_{(i,j)\in R_k \cup R_k^R} (\lambda_{S^f_{ij}} \cdot S^f_{ij}) + 
		   \mkern-20mu \displaystyle\sum_{(i,j)\in R_k \cup R_k^R} (\lambda_{V_{i}} \cdot V_{i}) + \\
		&  \frac{\rho}{2} [\mkern-20mu \displaystyle\sum_{(i,j)\in R_k \cup R_k^R} \lVert S^f_{ij} - S_{ij}^C + z_{S^f_{ij}}\rVert_2^2 + \nonumber\\
		  &  \mkern-1mu \displaystyle\sum_{(i,j)\in R_k \cup R_k^R} \lVert V_{i} - V_{i}^C + z_{V_{i}} \rVert_2^2 \;]  \nonumber\\		   
		\mbox{\bf subject to : }& \eqref{eq:reg_ac_1} - \eqref{eq:reg_ac_8} \nonumber
	\end{align}
	}
	\vspace{-10pt}
\end{model}

\begin{algorithm}[h]
\setcounter{AlgoLine}{0}
\SetKwInOut{Input}{Inputs}
\SetKwInOut{Output}{Output}
\caption{Cold-start: Two-level ADMM}
\label{alg:two-level_admm_init_cold}
{\footnotesize

\SetKwInOut{Output}{Output}
\SetKwProg{Fn}{Function}{:}{}
\SetKwProg{Init}{Initialize}{:}{}
\Fn{\texttt{initialize}(k)}{
  $\bm{S^C}[k] \gets (S^C_{ij} \gets 0 : (i,j) \in R_k \cup R_k^R)$\\ 
  $\bm{V^C}[k] \gets (V^C_{i} \gets 1 : (i,j) \in R_k \cup R_k^R)$\\
  $\bm{\lambda_S}[k] \gets ( \lambda_{S^f_{ij}} \gets 0 : (i,j) \in R_k \cup R_k^R)$ \\
  $\bm{\lambda_V}[k] \gets ( \lambda_{V_{i}} \gets 0 : (i,j) \in R_k \cup R_k^R)$\\
  $\bm{z_S}[k] \gets ( z_{S^f_{ij}} \gets 0 : (i,j) \in R_k \cup R_k^R)$\\
  $\bm{z_V}[k] \gets ( z_{V_{i}}    \gets 0 : (i,j) \in R_k \cup R_k^R)$\\
  $\bm{\Lambda_S}[k] \gets ( \Lambda_{S^f_{ij}} \gets 0 : (i,j) \in R_k \cup R_k^R)$\\
  $\bm{\Lambda_V}[k] \gets ( \Lambda_{V_{i}}    \gets 0 : (i,j) \in R_k \cup R_k^R)$\\  
}
}
\end{algorithm}

\begin{algorithm}[h]
\setcounter{AlgoLine}{0}
\SetKwInOut{Input}{Inputs}
\SetKwInOut{Output}{Output}
\caption{Warm-start with ML: Two-level ADMM}
\label{alg:two-level_admm_init_warm}
{\footnotesize

\SetKwInOut{Output}{Output}
\SetKwProg{Fn}{Function}{:}{}
\SetKwProg{Init}{Initialize}{:}{}
\Fn{\texttt{initialize}(k)}{
  $\bm{S^C}[k] \gets \hat{\bm{S}}^C[k] = (S^C_{ij} \gets \hat{S}^C_{ij} : (i,j) \in R_k \cup R_k^R)$\\ 
  $\bm{V^C}[k] \gets \hat{\bm{V}}^C[k] = (V^C_{i} \gets \hat{V}^C_{i} : (i,j) \in R_k \cup R_k^R)$\\
  $\bm{\lambda_S}[k] \gets \hat{\bm{\lambda}}_S[k] = ( \lambda_{S^f_{ij}} \gets \hat{\lambda}_{S^f_{ij}} : (i,j) \in R_k \cup R_k^R)$ \\
  $\bm{\lambda_V}[k] \gets \hat{\bm{\lambda}}_V[k] = ( \lambda_{V_{i}} \gets \hat{\lambda}_{V_{i}} : (i,j) \in R_k \cup R_k^R)$\\
  $\bm{z_S}[k] \gets \hat{\bm{z}}_S[k] = ( z_{S^f_{ij}} \gets \hat{z}_{S^f_{ij}} : (i,j) \in R_k \cup R_k^R)$ \\
  $\bm{z_V}[k] \gets \hat{\bm{z}}_V[k] = ( z_{V_{i}} \gets \hat{z}_{V_{i}} : (i,j) \in R_k \cup R_k^R)$\\
  $\bm{\Lambda_S}[k] \gets \hat{\bm{\Lambda}}_S[k] = ( \Lambda_{S^f_{ij}} \gets \hat{\Lambda}_{S^f_{ij}} : (i,j) \in R_k \cup R_k^R)$ \\
  $\bm{\Lambda_V}[k] \gets \hat{\bm{\Lambda}}_V[k] = ( \Lambda_{V_{i}} \gets \hat{\Lambda}_{V_{i}} : (i,j) \in R_k \cup R_k^R)$\\  
}
}
\end{algorithm}

\begin{algorithm}[h]
\SetKwInOut{Input}{Inputs}
\SetKwInOut{Output}{Output}
\caption{Two-level ADMM: Main routine}
\label{alg:two-level-admm}
{\footnotesize

\SetKwInOut{InputN}{Network data}
\SetKwInOut{InputA}{Search parameters}
\SetKwInOut{InputP}{Primal initial input}
\SetKwInOut{InputD}{Dual initial input}
\SetKwInOut{Output}{Output}
\SetKwFunction{PL}{$P_{L}$}
\SetKwProg{Fn}{Function}{:}{}
\SetKwProg{Mn}{Algorithm}{:}{}

\InputN{$\bm{\mathcal N}, \bf{S}^d$}
\InputA{$\rho_0, t_{max}$}
$\rho \gets \rho_0, \beta \gets 0.5 \rho_0$ \\
\For{$k \in K$}{
$\bm{S^C}[k], \bm{\lambda_S}[k], \bm{V^C}[k], \bm{\lambda_V}[k], 
\bm{z_S}[k], \bm{\Lambda_S}[k], \bm{z_V}[k], \bm{\Lambda_V}[k]$
$\gets 
\texttt{initialize}(k)
$ \\
}
\For{$t = 1, 2, \ldots, t_{max}$} 
{ 
 \For{$k \in K$}{ 
   Regional AC-OPF:\\
   $(S^f_{ij}, V_i: (i,j) \in R_k \cup R_k^R) \gets $  
   \PL{$\bm{S^d}[k], \bm{S^C}[k], \bm{\lambda_S}[k], \bm{z_S}[k], \bm{V^C}[k], \bm{\lambda_V}[k], \bm{z_V}[k] $} \\
   Slack parameter update:\\
   $\bm{z_S}[k] \gets (z_{S^f_{ij}} \gets (-\Lambda_{S^f_{ij}} - \lambda_{S^f_{ij}} - \rho(S^f_{ij} - S^C_{ij} )) / (\beta + \rho) : (i,j) \in R_k \cup R_k^R)$\\
   $\bm{z_V}[k] \gets (z_{V_{i}} \gets (-\Lambda_{V_{i}} - \lambda_{V_{i}} - \rho(V_{i} - V^C_{i} )) / (\beta + \rho) : (i,j) \in R_k \cup R_k^R)$\\
   Lagrange multiplier update:\\
   $\bm{\lambda_S}[k] \gets ( \lambda_{S^f_{ij}} \gets \lambda_{S^f_{ij}} + \rho(S^f_{ij} - S^C_{ij} + z_{S^f_{ij}}) : (i,j) \in R_k \cup R_k^R)$ \\
   $\bm{\lambda_V}[k] \gets ( \lambda_{V_{i}} \gets \lambda_{V_{i}} + \rho(V_{i} - V^C_{i} + z_{V_i}) : (i,j) \in R_k \cup R_k^R)$ \\
   Consensus update:\\
   $S^C_{ij}  \gets (S^C_{ij} + S^f_{ij}) / 2 : \forall (i,j) \in R_k \cup R_k^R$\\
   $V^C_{i}  \gets (V^C_{i} + V_i) / 2    : \forall (i,j) \in R_k \cup R_k^R$\\
 }
 \uIf{outer\_loop\_update\_criteria($t$)}{
    \For{$k \in K$}{
        \uIf{$\lVert vcat(\bm{z_S}[k], \bm{z_V}[k]) \rVert_2 \leq \eta(t, t_{max})$}{
        $\bm{\Lambda_S}[k] \gets \bm{\Lambda_S}[k] + \beta \bm{z_S}[k]$\\
        $\bm{\Lambda_V}[k] \gets \bm{\Lambda_V}[k] + \beta \bm{z_V}[k]$\\
        }
        \Else{
        $\beta \gets c_\beta \cdot \beta$\\
        $\rho \gets 2 \cdot \beta$\\
        }
    }
 }
}
}
\end{algorithm}

\begin{table}[h]
\caption{Two-level ADMM: Prediction Errors in \% for Various Filters. \label{tab:pred_error_two_level}}
\resizebox{0.99\linewidth}{!}
	{
	\begin{tabular}{| c | l | r r | r  r| r r| r r|}
    \toprule
    Network & Filter & $\bm{p}^C$ & $\bm{q}^C$ & $\bm{v}^C$ & $\bm{\theta}^C$ & 
                       $\lambda_{\bm{p}}$ & $\lambda_{\bm{q}}$ & $\lambda_{\bm{v}}$ & $\lambda_{\bm{\theta}}$ \\     
    \midrule
     \multirow{9}{*}{France\_EHV} & NIL & 8.55 & 16.93 & 5.44 & 9.01 & 5.17 & 71.52 & 68.41 & 39.29\\
     \cmidrule{2-10}
      & c(90\%) & 8.26 & 15.18 & 5.09 & 8.70 & 4.82 & 57.88 & 54.81 & 37.83\\
      & c(80\%) & 7.84 & 13.65 & 4.73 & 8.35 & 4.35 & 51.93 & 49.08 & 38.22\\
      & c(70\%) & 7.32 & 12.44 & 4.31 & 8.05 & 4.36 & 41.06 & 40.30 & 33.16\\
      & c(60\%) & 6.84 & 11.05 & 3.84 & 7.77 & 4.30 & 41.33 & 40.37 & 32.62\\
      & c(50\%) & \bf{6.36} & \bf{9.75} & \bf{3.37} & 7.15 & 4.02 & \bf{24.80} & \bf{25.71} & \bf{20.93}\\
    \cmidrule{2-10}
      & s(4.0) & 8.45 & 16.07 & 5.28 & 8.86 & 4.96 & 49.68 & 49.58 & 31.81\\
      & s(3.0) & 8.35 & 14.86 & 5.03 & 8.71 & 4.69 & 45.33 & 45.31 & 31.58\\
      & s(2.0) & 6.83 & 10.58 & 3.54 & \bf{7.00} & \bf{3.83} & 34.01 & 34.45 & 27.03\\
    \midrule
    \multirow{9}{*}{LYON}        & NIL & 10.68 & 22.22 & 7.85 & 7.13 & 11.40 & 39.16 & 38.60 & 45.80\\
    \cmidrule{2-10}
     & c(90\%) & 9.61 & 20.52 & 7.28 & 6.27 & 8.92 & 36.36 & 37.21 & 42.70\\
     & c(80\%) & 8.98 & 19.23 & 6.73 & 5.94 & 7.95 & 34.51 & 36.17 & 40.37\\
     & c(70\%) & 8.45 & 18.35 & 6.31 & 5.71 & 7.53 & 33.60 & 35.14 & 38.99\\
     & c(60\%) & 7.81 & 17.54 & 5.90 & 5.41 & 7.60 & 33.57 & 34.99 & 38.87\\
     & c(50\%) & \bf{7.20} & 16.57 & \bf{5.41} & 5.10 & 7.39 & 31.41 & 32.42 & 36.86\\
    \cmidrule{2-10}
     & s(4.0) & 9.93 & 20.99 & 7.48 & 6.44 & 9.27 & 37.01 & 37.33 & 44.30\\
     & s(3.0) & 9.07 & 19.51 & 6.89 & 5.86 & 7.84 & 34.35 & 34.94 & 41.40\\
     & s(2.0) & 7.25 & \bf{15.88} & 5.42 & \bf{4.65} & \bf{6.17} & \bf{28.85} & \bf{29.70} & \bf{32.59}\\                             
    \midrule
    \multirow{9}{*}{France}      & NIL & 12.87 & 22.13 & 7.71 & 16.16 & 6.01 & 71.94 & 78.65 & 38.50\\
    \cmidrule{2-10}
    & c(90\%) & 12.27 & 20.91 & 7.17 & 15.91 & 5.07 & 54.60 & 57.65 & 34.92\\
    & c(80\%) & 11.62 & 19.56 & 6.57 & 15.69 & 4.31 & 43.54 & 46.45 & 29.48\\
    & c(70\%) & 10.94 & 18.28 & 5.98 & 15.32 & 3.67 & 40.25 & 42.13 & 22.45\\
    & c(60\%) & 10.37 & 17.08 & 5.53 & 15.17 & 3.43 & 38.06 & 39.13 & 18.10\\
    & c(50\%) & 10.14 & 16.21 & \bf{5.30} & 15.22 & \bf{3.38} & \bf{35.52} & \bf{36.56} & \bf{17.64}\\
    \cmidrule{2-10}
    & s(4.0) & 12.58 & 21.55 & 7.46 & 15.90 & 5.62 & 64.77 & 71.81 & 37.70\\
    & s(3.0) & 11.89 & 20.73 & 7.06 & 15.42 & 5.13 & 59.59 & 64.65 & 35.98\\
    & s(2.0) & \bf{8.82} & \bf{15.29} & 5.42 & \bf{11.26} & 3.85 & 44.69 & 50.50 & 27.90\\                              
    \bottomrule
  	\end{tabular}
  	}
  	
\end{table}

\begin{table}[h]
\caption{Two-level ADMM: Mean Prediction Errors (in p.u.) for $\bm{z}$ with Various Filters. 
\label{tab:slack_pred_error_two_level}}
\centering
\resizebox{0.55\linewidth}{!}
	{
	\begin{tabular}{| c | l | r r | r  r|}
    \toprule
    Network & Filter & $\bm{z_p}$ & $\bm{z_q}$ & $\bm{z_v}$ & $\bm{z_\theta}$  \\     
    \midrule
     \multirow{9}{*}{France\_EHV} & NIL & 0.18 & 0.19 & 0.19 & 0.19\\
     \cmidrule{2-6}
      & c(90\%) & 0.18 & 0.18 & 0.18 & 0.19 \\
      & c(80\%) & 0.17 & 0.18 & 0.18 & 0.18 \\
      & c(70\%) & 0.17 & 0.17 & 0.17 & 0.17 \\
      & c(60\%) & 0.16 & 0.16 & 0.16 & 0.16 \\
      & c(50\%) & 0.15 & 0.16 & 0.16 & 0.16 \\
    \cmidrule{2-6}
      & s(4.0) & 0.18 & 0.19 & 0.19 & 0.19 \\
      & s(3.0) & 0.18 & 0.18 & 0.18 & 0.18 \\
      & s(2.0) & 0.15 & 0.15 & 0.16 & 0.16 \\
    \midrule
    \multirow{9}{*}{LYON}        & NIL & 0.27 & 0.27 & 0.28 & 0.30 \\
    \cmidrule{2-6}
     & c(90\%) & 0.26 & 0.26 & 0.27 & 0.28 \\
     & c(80\%) & 0.24 & 0.24 & 0.25 & 0.26 \\
     & c(70\%) & 0.23 & 0.23 & 0.24 & 0.25 \\
     & c(60\%) & 0.23 & 0.22 & 0.23 & 0.24 \\
     & c(50\%) & 0.22 & 0.22 & 0.22 & 0.23 \\
    \cmidrule{2-6}
     & s(4.0) & 0.26 & 0.26 & 0.27 & 0.28 \\
     & s(3.0) & 0.25 & 0.25 & 0.25 & 0.27 \\
     & s(2.0) & 0.22 & 0.22 & 0.23 & 0.24 \\                             
    \midrule
    \multirow{9}{*}{France}      & NIL & 0.33 & 0.33 & 0.34 & 0.32 \\
    \cmidrule{2-6}
    & c(90\%) & 0.31 & 0.31 & 0.32 & 0.31 \\
    & c(80\%) & 0.29 & 0.29 & 0.30 & 0.29  \\
    & c(70\%) & 0.27 & 0.28 & 0.28 & 0.27 \\
    & c(60\%) & 0.26 & 0.26 & 0.26 & 0.26 \\
    & c(50\%) & 0.25 & 0.25 & 0.26 & 0.25 \\
    \cmidrule{2-6}
    & s(4.0) & 0.32 & 0.32 & 0.33 & 0.32 \\
    & s(3.0) & 0.31 & 0.31 & 0.32 & 0.30 \\
    & s(2.0) & 0.32 & 0.31 & 0.31 & 0.32 \\                              
    \bottomrule
  	\end{tabular}
  	}  	
\end{table}

\begin{table}[h]
\caption{Two-level ADMM: Prediction Errors for $\Lambda$ (in \%) For Various Filters. 
\label{tab:dual_slack_pred_error_two_level}}
\centering
\resizebox{0.60\linewidth}{!}
	{
	\begin{tabular}{| c | l | r r | r  r|}
    \toprule
    Network & Filter & $\bm{\Lambda_p}$ & $\bm{\Lambda_q}$ & $\bm{\Lambda_v}$ & $\bm{\Lambda_\theta}$  \\     
    \midrule
     \multirow{9}{*}{France\_EHV} & NIL & 5.14 & 64.45 & 61.30 & 37.54 \\
     \cmidrule{2-6}
      & c(90\%) & 4.83 & 56.55 & 53.70 & 37.28  \\
      & c(80\%) & 4.34 & 50.71 & 48.36 & 37.57 \\
      & c(70\%) & 4.36 & 39.93 & 39.26 & 31.41 \\
      & c(60\%) & 4.28 & 40.05 & 39.44 & 30.79 \\
      & c(50\%) & 4.02 & 24.64 & 25.65 & 20.79 \\
    \cmidrule{2-6}
      & s(4.0) & 4.95 & 47.36 & 47.53 & 30.21 \\
      & s(3.0) & 4.69 & 44.00 & 44.15 & 30.67  \\
      & s(2.0) & 3.82 & 32.73 & 33.42 & 26.28 \\
    \midrule
    \multirow{9}{*}{LYON}        & NIL & 10.73 & 38.96 & 38.56 & 46.03 \\
    \cmidrule{2-6}
     & c(90\%) & 8.87 & 36.23 & 37.03 & 42.91 \\
     & c(80\%) & 7.97 & 34.40 & 35.94 & 40.57 \\
     & c(70\%) & 7.50 & 33.49 & 34.88 & 39.19 \\
     & c(60\%) & 7.51 & 33.45 & 34.72 & 39.08 \\
     & c(50\%) & 7.37 & 31.31 & 32.23 & 37.16 \\
    \cmidrule{2-6}
     & s(4.0) & 9.07 & 36.87 & 37.16 & 44.32 \\
     & s(3.0) & 7.81 & 34.31 & 34.79 & 41.63 \\
     & s(2.0) & 6.16 & 28.80 & 29.58 & 32.79 \\                             
    \midrule
    \multirow{9}{*}{France}      & NIL & 5.45 & 53.99 & 58.04 & 41.38 \\
    \cmidrule{2-6}
    & c(90\%) & 4.80 & 46.95 & 50.10 & 36.64 \\
    & c(80\%) & 4.24 & 42.91 & 45.57 & 30.23 \\
    & c(70\%) & 3.66 & 39.89 & 41.68 & 22.73 \\
    & c(60\%) & 3.44 & 37.87 & 38.79 & 18.26 \\
    & c(50\%) & 3.40 & 35.37 & 36.29 & 17.81 \\
    \cmidrule{2-6}
    & s(4.0) & 5.20 & 52.83 & 56.78 & 40.35 \\
    & s(3.0) & 4.84 & 49.82 & 54.08 & 38.43 \\
    & s(2.0) & 3.72 & 40.67 & 44.63 & 28.86 \\                              
    \bottomrule
  	\end{tabular}
  	}  	
\end{table}

\vfill

\end{document}